\documentclass[fleqn,usenatbib]{rasti}

\usepackage{newtxtext,newtxmath}

\usepackage[T1]{fontenc}

\DeclareRobustCommand{\VAN}[3]{#2}
\let\VANthebibliography\thebibliography
\def\thebibliography{\DeclareRobustCommand{\VAN}[3]{##3}\VANthebibliography}

\usepackage{graphicx}	
\usepackage{amsmath}	
\usepackage{multirow}
\usepackage{booktabs}
\usepackage{xspace}
\usepackage[
    frozencache,
    cachedir=_minted-cache
]{minted}
\usepackage{tikz}
\usetikzlibrary{shapes.geometric, arrows.meta, positioning, fit, calc}

\tikzset{
  startstop/.style = {rectangle, rounded corners, minimum width=3cm, minimum height=0.6cm, text centered, draw=black, fill=blue!40},
  process/.style   = {rectangle, minimum width=3cm, minimum height=0.6cm, text centered, draw=black, fill=blue!20},
  update-host/.style   = {rectangle, minimum width=3cm, minimum height=0.6cm, text centered, draw=black, fill=red!20},
    update-device/.style   = {rectangle, minimum width=3cm, minimum height=0.6cm, text centered, draw=black, fill=green!20},
  arrow/.style     = {thick, -Stealth},  
  hostprocess/.style   = {rectangle, minimum width=3cm, minimum height=0.6cm, text centered, draw=black, fill=orange!20},
}

\usepackage{caption}
\usepackage{subcaption}
\usepackage{longtable}

\newcommand{\amrvac}{\texttt{MPI-AMRVAC}\xspace}
\newcommand{\agile}{\texttt{AGILE}\xspace}
\newcommand{\bhac}{\texttt{BHAC}\xspace}
\newcommand{\gramrvac}{\texttt{GR-AMRVAC}\xspace}
\newcommand{\bhacplus}{\texttt{BHAC+}\xspace}
\newcommand{\gmunu}{\texttt{Gmunu}\xspace}
\newcommand{\bhacqgp}{\texttt{BHAC-QGP}\xspace}
\newcommand{\foap}{\texttt{foap4}\xspace}
\newcommand{\hamr}{\texttt{H-AMR}\xspace}
\newcommand{\athenapk}{\texttt{AthenaPK}\xspace}
\newcommand{\athenaK}{\texttt{AthenaK}\xspace}
\newcommand{\idefix}{\texttt{IDEFIX}\xspace}

\newcommand{\gamertwo}{\texttt{GAMER-2}\xspace}
\newcommand{\gramrx}{\texttt{GR-AMR-X}\xspace}
\newcommand{\gpluto}{\texttt{gPLUTO}\xspace}
\newcommand{\echo}{\texttt{ECHO}\xspace}
\newcommand{\batsrus}{\texttt{BATSRUS GPU}\xspace}
\newcommand{\mancha}{\texttt{Mancha3D}\xspace}

\title[AGILE 1.0]{Astrophysics on GPUs: introducing AGILE 1.0}

\author[O. Porth et al.]{Oliver Porth $^{1}$,\thanks{E-mail: o.porth@uva.nl (OP)}
Adrian Kelly$^{2}$, 
Olaf Willocx$^{2}$, 
Hao Wu$^{2,3}$, 
Jesse Vos$^{2}$, 
Yuhao Zhou$^{3}$, 
\newauthor
Héctor R. Olivares Sánchez$^{4}$, 
Leon Oostrum$^{5}$, 
Johan Hidding$^{5}$,
Victor Azizi$^{5}$, 
Chun Xia$^{6}$, 
\newauthor
Rony Keppens$^{2}$ 
and Jannis Teunissen$^{7,2}$ 
\\
$^{1}$Anton Pannekoek Institute for Astronomy, University of Amsterdam, Science Park 904, 1098 XH Amsterdam, The Netherlands\\
$^{2}$Centre for mathematical Plasma Astrophysics, Department of Mathematics,
KU Leuven, Celestijnenlaan 200B, B-3001 Leuven, Belgium\\
$^{3}$School of Astronomy and Space Science and Key Laboratory of Modern Astronomy and Astrophysics, Nanjing University, Nanjing, 210023, China\\
$^{4}$Departamento de Matemática da Universidade de Aveiro and
Centre for Research and Development in Mathematics and Applications (CIDMA),\\ Campus de Santiago, Aveiro, 3810-193, Portugal\\
$^{5}$Netherlands eScience Center, Science Park 402, Amsterdam, 1098 XH, The Netherlands\\
$^{6}$School of Physics and Astronomy, Yunnan University, Kunming 650500, China\\
$^{7}$Centrum Wiskunde \& Informatica, Science Park 123, Amsterdam, 1098 XG, The Netherlands
}

\date{Accepted XXX. Received YYY; in original form ZZZ}

\pubyear{\the\year{}}

\begin{document}
\label{firstpage}
\pagerange{\pageref{firstpage}--\pageref{lastpage}}
\maketitle

\begin{abstract}
We present \agile, a GPU-enabled adaptive mesh refinement (AMR) framework for the solution of (near-) conservation laws which occur in astro- and solar-physical applications.  
\agile  is written in modern fortran 2003, inherits a part of its modules and mesh handling from \amrvac, and achieves excellent GPU performance via OpenACC offloading.  We here discuss the design decisions which enable \agile to perform cost-efficient and scalable deeply nested AMR simulations with moderate block sizes of e.g. $16^3$ cells.  
\agile currently implements several physics modules, ie. hydrodynamics, frozen-field hydrodynamics, magnetohydrodynamics and special-relativistic hydrodynamics and can easily be extended further through its modular design.  
Besides strong scaling tests to up to 2048 GPUs and standard benchmarks which show consistent performance across a large range of devices and problem sizes, we demonstrate \agile's capabilities by means of state-of-the art science applications with all currently available physics modules.  
\end{abstract}




\section{Introduction}

Simulations of hydro- (HD) and magneto-hydrodynamical (MHD) flows are an essential component of theoretical astrophysics and modeling.  Astrophysical problems are often characterized by extreme dynamic range and turbulence, translating to the need for large-scale simulations to obtain the appropriate separation of temporal and spatial scales.  
Fortunately, the relevant compressible fluid dynamics is governed by conservation laws that are amenable to explicit integration algorithms which are easy to parallelize and often show ideal scaling even when using entire supercomputer facilities. 
In recent years, due to their increased energy efficiency compared to central processing units (CPUs), performance on supercomputers is mostly obtained through the use of accelerator cards, primarily graphics processing units (GPUs).  
To benefit from increasing compute capabilities, it is hence essential that astrophysical codes are GPU enabled and existing algorithms are optimized for the hardware specific to GPU clusters.  

In this paper, we introduce \agile, a new adaptive-mesh-refinement (AMR) framework to solve near-conservation laws on GPUs.  \agile is based on \amrvac \citep{amrvac2012,PorthXia2014,XiaTeunissenEtAl2018a,Jannis2019,KeppensTeunissenEtAl2021,KeppensPopescuBraileanuEtAl2023}, to which it remains largely compatible.  
The decision to keep a high level of compatibility with \amrvac and not undertake a full re-write is based on the large user- and knowledge-base of this public code\footnote{\url{http://www.amrvac.org}} which has initiated several spin-off projects such as the general relativistic codes \bhac\footnote{\url{http://bhac.science}} \citep{PorthOlivares2017,OlivaresPorthEtAl2019a}, \gramrvac \citep{CasseVarniereEtAl2017}, \gmunu \citep{CheongLinEtAl2020}, \bhacplus \citep{NgJiangEtAl2024} (the former two with fixed metrics, the latter two with approximate spacetime evolution) and the heavy-ion-collision code \bhacqgp \citep{MayerDashEtAl2024}.  While these codes are primarily focused on astrophysical applications, the capability to efficiently solve partial differential equations can lead to much broader inter-disciplinary applications \citep{KeppensTeunissenEtAl2021}.  

The philosophy of \agile (``Astrophysics on GPUs for InterdiscipLinary Exascale challenges'') is hence to provide a flexible framework for the solution of (near-) conservation laws with primary focus on astrophysics, but not strictly limited to it.  
The algorithms and datastructures used in \agile are informed by our experiences with \foap, a minimal, yet fully featured AMR application that was built from scratch within our collaboration \citep[][]{TeunissenOlivaresSanchezEtAl2026}.  

\agile is neither the first nor the last GPU-enabled framework for astrophysical (magneto-) hydrodynamics, in fact recent years have seen a large increase of more or less specialized astrophysical codes being ported to GPUs.  Some examples are \hamr \citep{LiskaChatterjeeEtAl2022}, \athenapk \citep{GreteDolenceEtAl2022}, \gramrx \citep{ShankarMostaEtAl2023}, \athenaK \citep{StoneMullenEtAl2026}, \idefix \citep{LesurBaghdadiEtAl2023}, \gamertwo \citep{SchiveZuHoneEtAl2018}, \gpluto \citep{RossazzaMignoneEtAl2025}, \echo \citep{DelZannaLandiEtAl2024} \mancha \citep{ModestovAsensioEtAl2026} among others.  
By building \agile, our ambition is that users trained with \amrvac will be able to seamlessly transition to use GPU equipped clusters.  

This paper is organized as follows: In Section \ref{sec:design}, we elaborate on the design decisions underlying \agile and introduce the physics modules that are already implemented.  In Section \ref{sec:performance}, we demonstrate the performance and scaling.  In Section \ref{sec:examples}, we present various example applications with each physics module.  Section \ref{sec:conclusions} concludes and gives an outlook on planned future developments.

\section{Computational infrastructure}\label{sec:design}

\subsection{Design decisions} 

\subsubsection{Offloading framework}
There are many paradigms for GPU programming: vendor specific low-level kernel languages like Cuda and HIP,  performance portability layers such as Kokkos\footnote{\url{https://kokkos.org/}}, SYCL\footnote{\url{https://www.khronos.org/sycl/}}, Alpaka\footnote{\url{https://github.com/alpaka-group/alpaka}} and RAJA\footnote{\url{https://github.com/LLNL/RAJA}} which are mostly designed for the C++ ecosystem, language intrinsic parallel constructs such as fortran's \texttt{do concurrent} and directive based approaches such as OpenMP\footnote{\url{https://www.OpenMP.org/}} and OpenACC\footnote{\url{https://www.OpenACC.org/}}.  
A comprehensive overview of these approaches together with discussion of their performance portability can be found for example in \cite{DavisSivaramanEtAl2024}.  

To enable GPU acceleration for large (fortran) code-bases like \amrvac ($\approx 100\, 000$ lines of code), the directive based approaches are particularly attractive: simply by adding a few annotations, entire code blocks (loops) of the native language can be compiled into device code. 
OpenMP and OpenACC also enable explicit control of data transfer between host and device, a feature currently missing in the more generic language intrinsics. 

Meanwhile, the OpenACC model -- in particular when combined with the Nvidia HPC toolkit -- is quite mature and can reach performance on-par with native Cuda implementations \citep{HoshinoMaruyamaEtAl2013,OyarzunMiraEtAl2021,ShanAraya-Polo2024}.  It is encouraging that several production-grade computational fluid dynamics codes have successfully applied OpenACC offloading and have already demonstrated up to Exascale performance \citep{VincentGongEtAl2022, AnChenEtAl2025, RossazzaMignoneEtAl2025, WilfongLeBerreEtAl2026}.  

We therefore base \agile's offloading strategy on OpenACC, which achieves performance portability via parallelizing compilers that build the same source code for a variety of accelerator targets such as GPUs and many-processor architectures.  Current free compilers with OpenACC support are for example the Nvidia compiler suite, LLVM-flang/clang, and gcc.  Furthermore, we have developed \agile to run on the commercial HPE/Cray compiler which enables offloading to AMD cards on Cray HPC systems.

While OpenACC's high level approach allows to utilize accelerators with moderate programming effort, one downside is the lack of fine-grain control, in particular a missing programming interface for the GPU shared memory.  This aspect requires some attention by the programmer as the compiler must be able to determine whether the data will fit into the fast local memory at compile time, otherwise significant performance degradation can occur.  

\subsubsection{Build system and code structure}
\agile's build system has been significantly redesigned compared to \amrvac (which builds the entire code-base into a single library): to allow kernel optimizations at compile time, \agile generates specific binaries for each selected use case.  
The required options are stored in a single place: the simulation parameter file (\texttt{agile.par}) which is read in a pre-compilation step using the f90nml namelist parser in python.  Build dependencies are then computed on-the-fly using the fortdep\footnote{\url{https://github.com/gronki/fortdep}} tool.  To avoid frequent recompilations, each build is identified by a hash that allows to re-use the build folder for subsequent builds with the same options.  

Another major change from \amrvac concerns the macro language. \amrvac makes extensive use of the LASY preprocessor language \citep{Toth1997} which is used to translate dimensionally independent source into native 1D, 2D or 3D fortran code.  On the one hand, being able to compile the same source for multi-D applications is one of the strengths of \amrvac: the user can rapidly develop and test 2D setups before going into 3D production.  On the other hand, the idiosyncrasies of the LASY syntax pose a significant entry barrier, in particular for experienced programmers and AI assisted coding.  
Typically, 3D production runs are significantly more costly than 2D setups and will therefore benefit most from GPU acceleration. 
We expect that for 1D and 2D scenarios, a user might instead employ  \amrvac or \bhac which will remain relevant for this use-case on CPUs.  
For these reasons, \agile restricts to 3D-only setups and drops the LASY syntax legacy.  Instead, \agile uses the more widely supported fypp\footnote{\url{https://github.com/aradi/fypp}} as a pre-compilation language which is used for templating and code instantiation for efficient inlining across source files.  

\begin{figure}
  \centering
  \resizebox{0.5\textwidth}{!}{%
\begin{tikzpicture}[
  every node/.style={draw, rectangle, align=center}]

\node[hostprocess] (start) {Init level one};

\node[hostprocess, right of=start, node distance=4cm] (alloc1)
  {Allocate AMR node};

\node[hostprocess, below of=alloc1, node distance=1cm] (ic1) {Initial condition};

\node[update-device, right of=ic1, node distance=4cm] (upddev1)
  {\texttt{update device(w)}};

\draw[arrow] (start.east)   -- (alloc1.west);
\draw[arrow] (start.east) -- ++(0.5cm,0) |- (ic1.west);       
\draw[arrow] (ic1.east)   -- (upddev1.west);
  
\node(errest1) [process, below of=start, node distance=2cm] {AMR error estimator};
\node[update-host, right of=errest1, node distance=4.6cm] (up1) {\texttt{update host(refine,coarsen)}};
\draw[arrow] (errest1.east)   -- (up1.west);

\node(amr1) [hostprocess, below of=errest1, node distance=1cm] {AMR Coarsen Refine} ;
\node(alloc2) [hostprocess, right of=amr1, node distance=4cm]{Allocate AMR nodes};
\node[hostprocess, below of=alloc2, node distance=1cm] (ic2) {Initial condition};
\node[update-device, right of=ic2, node distance=4cm] (up2) {\texttt{update device(w)}};

\node(loadbalance1) [process, below of=ic2, node distance=1cm] {Load balance};
\node(connectivity1) [hostprocess, below of=loadbalance1, node distance=1cm] {Rebuild connectivity};
\node[update-device, right of=connectivity1, node distance=4cm] (nbinfo1) {\texttt{update device(nbinfo)}};
\node(bc0) [process, below of=connectivity1, node distance=1cm] {Boundary Conditions};

\draw[arrow] (amr1.east) -- (alloc2.west);
\draw[arrow] (amr1.east) -- ++(0.5cm,0) |- (ic2.west);       
\draw[arrow] (ic2.east) -- (up2.west);
\draw[arrow] (amr1.east) -- ++(0.5cm,0) |- (loadbalance1.west);       
\draw[arrow] (amr1.east) -- ++(0.5cm,0) |- (connectivity1.west);       
\draw[arrow] (connectivity1.east) -- (nbinfo1.west);
\draw[arrow] (amr1.east) -- ++(0.5cm,0) |- (bc0.west);       
\node[
  draw,
  rounded corners,
  dashed,
  inner sep=8pt,
  fit=(start) (amr1) (alloc2) (ic2) (up2) (bc0),
  label={[font=\large]above:Initial condition}
] (initbox) {};

\draw[arrow]
  (bc0.south) -- ++(0,-5mm) coordinate (p1)           
  -- ($ (p1) + (-6cm,0) $) coordinate (p2)            
  -- (p2 |- errest1.west)                               
  -- (errest1.west);                                    

\end{tikzpicture}
}
\caption{Flow chart of the initialization using host routines. 
The left column indicates subroutines called from the top-level  function, subsequent subroutine calls are shown correspondingly to the right.  
Data transfers from host to device are colored green, the opposite direction is colored red.  Host routines are marked orange and routines that execute GPU kernels are colored blue.
First the base level grid is initialized, an error estimator is employed on the device and flags to either coarsen or refine a block are updated on the host.  Refinement of the tree structure and allocation of new block metadata is done on the host.  The initial condition is again used to initialize newly generated blocks.  The ``AMR error estimator'' part is repeated up to \texttt{refine\_max\_level}-1 times, where \texttt{refine\_max\_level} is the user-specified maximum AMR level.  }
\label{fig:flowchart-init}
\end{figure}
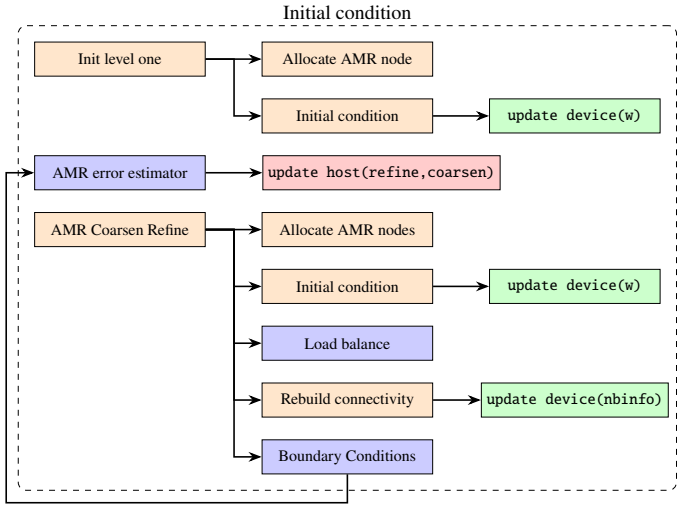

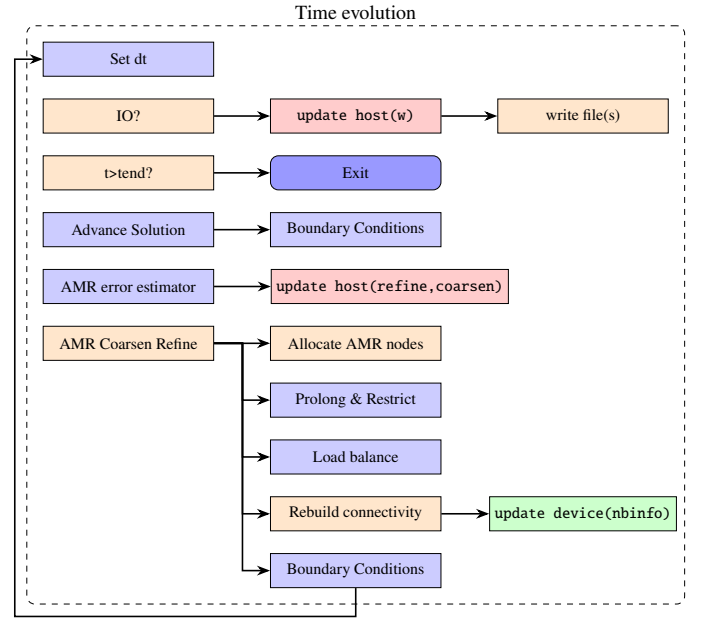
\begin{figure}
  \centering
  \resizebox{0.5\textwidth}{!}{%
\begin{tikzpicture}[
  every node/.style={draw, rectangle, align=center}]
\node(setdt) [process, below of=amr1, node distance=5.5cm] {Set dt};
\node(io)   [hostprocess, below of=setdt, node distance=1cm]  {IO?};
\node(up3)  [update-host, right of=io, node distance=4cm] {\texttt{update host(w)}};
\draw[arrow] (io.east) -- (up3.west);
\node(write)[hostprocess, right of=up3, node distance=4cm] {write file(s)};
\draw[arrow] (up3) -- (write.west);

\node(tend) [hostprocess, below of=io, node distance=1cm] {t>tend?};
\node(exit) [startstop, right of=tend, node distance=4cm]{Exit};
\draw[arrow] (tend.east) -- (exit.west);

\node(advance) [process, below of=tend, node distance=1cm] {Advance Solution};
\node(bc1) [process, right of=advance, node distance=4cm] {Boundary Conditions};
\draw[arrow] (advance.east) -- (bc1.west);

\node(errest2) [process, below of=advance, node distance=1cm] {AMR error estimator};
\node(up4) [update-host, right of=errest2, node distance=4.6cm] {\texttt{update host(refine,coarsen)}};
\draw[arrow] (errest2.east) -- (up4.west);

\node(amr2) [hostprocess, below of=errest2, node distance=1cm] {AMR Coarsen Refine} ;
\node(alloc2) [hostprocess, right of=amr2, node distance=4cm] {Allocate AMR nodes};
\node(prolong) [process, below of=alloc2, node distance=1cm] {Prolong \& Restrict};
\node(loadbalance) [process, below of=prolong, node distance=1cm] {Load balance};
\node(connectivity) [hostprocess, below of=loadbalance, node distance=1cm] {Rebuild connectivity};
\node[update-device, right of=connectivity, node distance=4cm] (nbinfo2) {\texttt{update device(nbinfo)}};
\node(bc2) [process, below of=connectivity, node distance=1cm] {Boundary Conditions};
\draw[arrow] (amr2.east) -- (alloc2.west);
\draw[arrow] (amr2.east) -- ++(0.5cm,0) |- (prolong.west);       
\draw[arrow] (amr2.east) -- ++(0.5cm,0) |- (loadbalance.west);       
\draw[arrow] (amr2.east) -- ++(0.5cm,0) |- (connectivity.west);       
\draw[arrow] (connectivity.east) -- (nbinfo2.west);
\draw[arrow] (amr2.east) -- ++(0.5cm,0) |- (bc2.west);       

\draw[arrow]
  (bc2.south) -- ++(0,-5mm) coordinate (p1)           
  -- ($ (p1) + (-6cm,0) $) coordinate (p2)            
  -- (p2 |- setdt.west)                               
  -- (setdt.west);                                    

\node[
  draw,
  rounded corners,
  dashed,
  inner sep=8pt,
  fit=(setdt) (bc2) (write),
  label={[font=\large]above:Time evolution}
] (initbox) {};

\end{tikzpicture}
}
\caption{Flow chart of the time evolution using device routines for operations on data and host routines for book keeping and IO.  Color scheme as in Figure \ref{fig:flowchart-init}.}
\label{fig:flowchart-time}
\end{figure}

Since there are many science ready setups available and postprocessing workflows are well established, \agile strives to be compatible with \amrvac as much as possible on these aspects.  Therefore, in this version 1.0, the initial conditions are computed by the same host subroutines (specified by the user in \texttt{mod\_usr.fpp}) and all IO occurs through the well tested host routines. After the initialization of the grid, data is transferred to the device and the computation proceeds fully device resident until the next IO event.  Thus everything related to data on the computational grid such as the time-update of the solution, AMR error estimation, prolongation/restriction operations, time step calculation, ghost-cell exchange etc. proceeds through device kernels without requiring any host/device transfer of mesh data.  
On the other hand, book-keeping such as keeping track of the grid connectivity and MPI neighbor relations is done via host routines.  These operations are essentially serial and do not benefit from GPU processing.  A flow chart of the initialization stage on the host is shown in Figure \ref{fig:flowchart-init}, and a corresponding chart for the time evolution is given in Figure \ref{fig:flowchart-time}.  

\subsubsection{Performance critical elements}

At the heart of \agile is the loop that updates the computational cells by one sub-step.  We have designed this part following our findings from the experimental \foap code \cite{TeunissenOlivaresSanchezEtAl2026}.  The algorithm is sketched in Listing \ref{lst:main-loop}.  Conceptually, \textit{\agile launches a single macro-kernel} to loop over the space-filling-curve of blocks via a \texttt{gang}-loop, followed by collapsed \texttt{vector}-loops over the cells.  On GPUs, operations for each block are thus kept local to one Streaming Multiprocessor (SMP).  Each vector unit within the SMP operates on a single cell to apply the spatial reconstruction and compute/apply fluxes  as well as source terms if requested by the physics module (cell-local and non-local e.g. gradient sources).  The main update kernel is sketched in Figure \ref{lst:main-loop}.

\begin{figure}
  \begin{minted}{fortran}
!$acc parallel loop gang 
do n = 1, nblocks
  !$acc loop vector collapse(3)
   do k = kmin, kmax
     do j = jmin, jmax
       do i = imin, imax
         ! Operate on u(1:nvars) = w(i,j,k,1:nvars,n):
         ! 1. Apply fluxes in all dimensions
         ! 2. Apply local source terms
         ! 3. Apply non-local source terms
      end do
    end do
  end do
end do
\end{minted}
\caption{Main integration kernel looping over blocks via OpenACCs ``gang loop'' and scheduling operations on a cell-by-cell basis on each vector processor.  }
\label{lst:main-loop}
\end{figure}

The macro-kernel approach minimizes overhead from kernel launching as the entire grid is treated within a single kernel.  
The cell-local approach minimizes temporary shared memory usage since only single-cell values are created which can be kept on the registers.  This is in contrast to the philosophy of \amrvac where operations are always scheduled for entire blocks, leading to the creation of temporary outputs of the corresponding size.  

It is interesting to note that since each cell is bounded by two interfaces (in each direction), this algorithm performs redundant computations of the fluxes.  That is, the ``left'' interface fluxes of cell $(i+1,j,k)$ have already been computed as ``right'' fluxes by the thread working on cell $(i,j,k)$.  However, on massively parallel GPUs, we find that it is computationally advantageous to perform the extra calculations compared to sharing data across threads via a block-sized array.  
In section \ref{sec:singleGPU}, by varying the number of blocks and cells within a block, we will discuss how the performance of this algorithm is affected by shifting load between the outer gang-loop and the inner vector loops.   

Next to the device data handling and the main update kernel, another performance critical element is the ghostcell exchange between MPI tasks.  
Efficient buffer storage management is important since GPU memory is limited, in fact our first naive approach (allocate $(N_{\rm ranks}-1) \times N_{\rm max\_blocks} \times {\rm size\_per\_block}$) led to prohibitive memory overheads already at 64 ranks.  

To minimize the number of messages between neighboring processors, for each neighbor relation (neighbor on same-resolution level ``\texttt{srl}'', finer ``\texttt{f}'' or coarser ``\texttt{c}''), \agile packs all data to be communicated to a specific neighboring processor into a single buffer which is then exchanged by a single non-blocking \texttt{MPI\_Isend/MPI\_Irecv} pair.  Internally the bookkeeping is done by means of nested lists which allows \agile to provision exactly the amount of buffer space that is required.  
For example, for the same-resolution level neighbor type, the block indices with data to be sent to neighbor number \texttt{inb} are found in \texttt{nbinfo\%srl(inb)\%igrid(1:nigrids)} and the corresponding buffer is \texttt{nbinfo\%srl\_send(inb)\%buffer(:)} which holds the values of all exchanged cells in a flattened array.  These lists are updated upon mesh refinement events along with the structures for grid-connectivity on the host.  
To simplify the interpretation of the received data, along with the ghostcell data itself, we also send a flat integer array which provides the destination indices as seen from the receiving end. For example, in the srl case, we send the local ``\texttt{igrid}'' and the local specifiers of the boundary in question encoded via $\texttt{i1,i2,i3}\in [-1,0,1]$.   Here e.g. $[\texttt{i1,i2,i3,igrid}]=[-1,0,0,666]$ would specify the ``left'' face in x-direction as seen from the grid block with index 666.  
This approach turns out similar to \batsrus \citep[see][and the discussion therein]{AnChenEtAl2025} which also exchanges ``memory maps'' along with the buffers.  An alternative is presented in \foap \citep{TeunissenFOAP} which instead sorts received data according to a global order.  The clear advantage is that only essential data is communicated, the price to pay is a more complex algorithm which is less easily extensible e.g. to the case where only specific directions or data between certain AMR levels should be exchanged.  

In our experience, not all OpenACC implementations reliably support to dynamically resize and update nested structure members on the device. We hence resort to deleting and re-creating the entire \texttt{nbinfo} structure each time the grid changes.  To minimize the data transfer overheads, we found that it is advantageous to keep the number of OpenACC API calls to a minimum, meaning we combine OpenACC ``copyin'' calls for \texttt{nbinfo} members wherever possible.  Another option would be to use ``deepcopy'' which is however not uniformly supported across implementations.    Fortunately, when combining API calls, the cost of the ``copyin'' operation for the \texttt{nbinfo} structure is small compared to the finite volume macro-kernel and becomes negligible when multi-step RK methods are used.  

Buffer packing and unpacking happens on the device and the data is exchanged by direct memory access (GPUDirect RDMA for Nvidia and DirectGMA for AMD) between devices as signalled by \texttt{!\$acc host data use\_device()} pragmas.  If however RDMA is not available on a given hardware, \agile can be configured to always communicate updated host data instead via the compilation flag \texttt{make NOGPUDIRECT=1}.  

\subsection{Software development policy}
The development of the \agile framework itself takes place on the public repository \url{https://github.com/amrvac/AGILE} and \agile is released under the GPLv3 licence. Before addition of a substantial new feature, developers are advised to provide a test case which is automatically executed by the github continuous integration for each pull request.  At present, we build and auto-test \agile on three CPU target platforms: ubuntu/Intel OneAPI, ubuntu/gcc, macOS/gcc through github runners.  
In addition, a custom github runner builds and executes \agile on an university-hosted workstation featuring dual Nvidia L40S GPUs.  
To further improve code integrity, the master branch can only be updated after review via pull requests. 
Application codes can be developed quite independently and privately as physics modules due to the modular structure of \agile.

\subsection{Physics modules}

\agile is designed to solve (near-) conservation laws of the general form
\begin{equation}
    \partial_t \mathbf{U} + \nabla \cdot \mathbf{F}(\mathbf{U}) = \mathbf{S}(\mathbf{U},\nabla \mathbf{U},\mathbf{x},t)
\end{equation}
which comprises numerous astrophysically relevant partial differential equations such as Newtonian and General relativistic (magneto-) hydrodynamics.  
\agile is fully modular and exposes generic interfaces of fluxes and source terms to the update algorithm.  Specific implementations are collected into physics modules and corresponding GPU kernels are assembled during the build process.  This allows the compiler to build maximally optimized kernels, while the source maintains a modular structure.  

\agile's physics subroutines operate on a cell-by-cell basis.  This results in compact intuitively understandable code in the physics modules.  To give an example, the \texttt{get\_flux()} subroutine for Newtonian hydrodynamics is shown in Figure \ref{lst:getflux}.  

\begin{figure}
  \begin{minted}{fortran}
subroutine get_flux(u, xC, flux_dim, flux)
  !$acc routine seq
  real(dp), intent(in)  :: u(nw_phys)
  real(dp), intent(in)  :: xC(ndim)
  integer, intent(in)   :: flux_dim
  real(dp), intent(out) :: flux(nw_flux)

  ! Density flux
  flux(iw_rho) = u(iw_rho) * u(iw_mom(flux_dim))

  ! Momentum flux
  flux(iw_mom(1)) = u(iw_rho) * u(iw_mom(1)) &
  * u(iw_mom(flux_dim))
  flux(iw_mom(2)) = u(iw_rho) * u(iw_mom(2)) &
  * u(iw_mom(flux_dim))
  flux(iw_mom(3)) = u(iw_rho) * u(iw_mom(3)) &
  * u(iw_mom(flux_dim))

  flux(iw_mom(flux_dim)) = flux(iw_mom(flux_dim)) & 
  + u(iw_e)

  ! Energy flux
  flux(iw_e) = u(iw_mom(flux_dim)) &
  * (u(iw_e)/(hd_gamma - 1.0_dp) + 0.5_dp * &
     u(iw_rho) * sum(u(iw_mom(1:ndim))**2) + u(iw_e))

end subroutine get_flux
\end{minted}
\caption{Device routine to obtain the hydrodynamic fluxes from variables in primitive form in \texttt{u} (hence e.g. \texttt{u(iw\_e)} contains pressure etc.).  Note that a developer does not need to add parallelization instructions since the routine is executed on the vector level.}
\label{lst:getflux}
\end{figure}
Note that besides the \texttt{!\$acc routine seq} ~pragma, no GPU specific code is needed as the physics modules are embedded into the update algorithm on the vector-level.  In our experience, porting of an existing physics module from \amrvac to \agile can be done in a timeframe of a day.  

Next to specific physics modules, \agile also provides functionality that can be used across physics types, namely the radiative cooling and gravity source term modules.  
In addition, a user can freely specify boundary conditions, user-defined source terms and custom AMR triggers in \agile's \texttt{mod\_usr.fpp} source file.  

In the following sections, we briefly go over the physics modules already available in \agile; a corresponding overview is given in Table \ref{tab:physics}.  

\subsubsection{Hydrodynamics} \label{sec:hd}
The hydrodynamic module solves the partial differential equations (PDEs) that govern compressible gas dynamics. We work with conservative variables consisting of density, vectorial momentum density, hydrodynamic energy density and a user-selected set of $n_\mathrm{tr}$ tracers collected in $\mathbf{U}=[\rho,\mathbf{m}\equiv\rho\mathbf{v},e_{\mathrm{hd}},D_{\mathrm tr}^{i=1,n_{\mathrm tr}}]$ that obey
\begin{eqnarray}
 \partial_t \rho + \nabla \cdot (\rho\mathbf{v}) & =  & 0  \,, \label{q-mass}\\
  \partial_t \mathbf{m} + \nabla \cdot \left(\mathbf{v}\mathbf{m}+p\mathbf{I}\right) & = & \rho\mathbf{g} \,, \label{q-mom}\\
  \partial_t e_{\mathrm{hd}}+\nabla\cdot\left[\mathbf{v}(e_{\mathrm{hd}}+p)\right]
     & = &
     \rho\mathbf{v}\cdot\mathbf{g} -n_e  n_H\Lambda(T) \,, \label{q-ene}\\
  \partial_t D_{tr}^i+\nabla \cdot(\mathbf{v} D_{tr}^i) & = & 0 \,. \label{q-tr}
 \end{eqnarray}
Here $p$ is the gas pressure, $\mathbf{v}$ the velocity vector and $\mathbf{I}$ the identity tensor. Tracers are handled using their conservative variable $D_{tr}=\rho\vartheta_{tr}$, such that the actual tracer value $\vartheta_{tr}$ obeys a pure advection equation.
Pre-implemented sources on the right-hand-side (RHS) allow the user to set the external gravitational acceleration $\mathbf{g}$ and select the treatment of local and instantaneous radiative losses as appropriate for optically thin radiative conditions. The numerical handling of the optically thin radiative sink term uses the exact integration method from \cite{Townsend2009}. This involves a tabulated radiative loss function $\Lambda(T)$ that only depends on the local temperature $T$. These losses in essence scale with the squared density, whereby $\rho$ is converted to electron $n_e$ and Hydrogen $n_H$ number densities under the assumption of constant and fully ionized conditions using a fixed Helium abundance $\mathrm{He}_a\equiv n_{He}/n_H$, where charge neutrality then sets $n_e=n_H(1+2 {\mathrm{He_a}})$ and $\rho=n_Hm_p(1+4\mathrm{He}_a)$ for proton mass $m_p$. Future work could generalize this prescription to spatio-temporally varying ionization degrees. The same is true for the hydrodynamic energy density and the corresponding closure for the internal energy density and pressure, which is currently set by
\begin{equation}
    e_{\mathrm{hd}}=\frac{p}{\hat{\gamma}-1}+\frac{1}{2}\rho v^2 \,,
\end{equation}
introducing an equation parameter $\hat{\gamma}$ representing the constant ratio of specific heats, i.e. $5/3$ for a mono-atomic gas. Finally, an ideal gas law relates $p={\cal{R}}\rho T$, with a fixed gas constant ${\cal{R}}=k_B/(\tilde{\mu}m_p)$ where $k_B$ denotes the Boltzmann constant. Consistent with the number density factors above, we adopt the constant mean molecular weight $\tilde{\mu}=(1+4\mathrm{He}_a)/(2+3\mathrm{He}_a)$. 

\subsubsection{Frozen-field hydrodynamics} \label{sec:ffhd}
As an intermediate towards full 3D magnetohydrodynamic (MHD) applications, we also ported the frozen-field hydrodynamic module that was introduced recently by \cite{Zhou2024,Zhou2025} to \agile. In a given {\em frozen} 3D magnetic topology $\mathbf{B}(\mathbf{x})$ that does not evolve in time, this module can solve for the set $\mathbf{U}=[\rho, m_{\parallel}=\rho v_{\parallel},e_{hd\parallel},q_\parallel]$ which obey mass conservation from Eq.~(\ref{q-mass}), but where we only deal with the magnetic field-line projected variant of Eq.~(\ref{q-mom}). Introducing the local magnetic unit vector $\mathbf{\hat{b}}=\mathbf{B}/B$ this means we solve for
\begin{eqnarray}
    \partial_t \rho+\nabla \cdot  \left (  \rho v_{\parallel} \mathbf{\hat{b}}\right ) &=& 0,
    \label{q-massff}\\
    \partial_t m_{\parallel}+\nabla \cdot \left [  \left (  \rho v^2_{\parallel} +p\right ){\mathbf{\hat{b}}}\right ] &=& \rho \mathbf{g}\cdot\mathbf{\hat{b}} + p (\nabla\cdot \mathbf{\hat{b}}),
    \label{q-momff}\\
    \partial_t e_{hd\parallel} + \nabla \cdot \left[\left((e_{hd\parallel} + p) v_{\parallel}-q_\parallel\right) \mathbf{\hat{b}}\right] &=& \rho v_{\parallel} \mathbf{g}\cdot\mathbf{\hat{b}}  -n_e  n_H\Lambda(T) \nonumber \\
    & & + H_{user}(\mathbf{x})\,,\label{q-enff} \\
    \partial_t q_\parallel & = & -\frac{\left(q_\parallel-\kappa_\parallel\mathbf{\hat{b}}\cdot \nabla T\right)}{\tau} \,, \nonumber \\
    & & \label{q-htc}
\end{eqnarray}
where now $e_{hd\parallel}=p/(\hat{\gamma}-1)+\rho v_\parallel^2/2$.
This uses a hyperbolic approach to handle the anisotropic (purely field-aligned) thermal conduction, where $\kappa_\parallel=\kappa_{\parallel,0}T^{5/2}$. This prescription forces the parallel heat flux $q_\parallel$ to approach its target expression $\kappa_\parallel \mathbf{\hat{b}}\cdot \nabla T$ within at most four CFL limited timesteps (hence setting $\tau=\max(4\Delta t_n, \tau_{\rm reduced})$, where various prescriptions for $\tau_{\rm reduced}$ are possible) \citep{Rempel2017,Warnecke2020,Navarro2022,Zhou2025}.
A purely local user-defined heating prescription can be coded up by the user in $H_{user}(\mathbf{x})$. Note that we re-use the external gravity module from the hydro case, but here only effective gravity along the fixed field enters. The radiative loss treatment is shared with all other physics modules.
\subsubsection{Magnetohydrodynamics} \label{sec:mhd}
The MHD module currently combines mass conservation from Eq.~(\ref{q-mass}), the ability for tracers from Eq.~(\ref{q-tr}), with the following generalizations of momentum and energy
\begin{equation}
\partial_t \mathbf{m} + \nabla \cdot \left(\mathbf{v}\mathbf{m}+(p+{\scriptstyle{\frac{1}{2}}}B^2)\mathbf{I}-\mathbf{B}\mathbf{B}\right)  = \rho\mathbf{g} 
\end{equation}

\begin{multline}
  \partial_t e_{\mathrm{mhd}}+\nabla\cdot\left[\mathbf{v}(e_{\mathrm{mhd}}+p+{\scriptstyle{\frac{1}{2}}}B^2)-\mathbf{B}(\mathbf{v}\cdot\mathbf{B})\right] \\
  \, \, = \rho\mathbf{v}\cdot\mathbf{g} -n_e  n_H\Lambda(T) + \nabla \cdot \left[ \mathbf{B} \times \eta \mathbf{J}\right]\,, \\
   \mathrm{where\,\,\,}  e_{\mathrm{mhd}}=\frac{p}{\hat{\gamma}-1}+\frac{1}{2}\rho v^2+\frac{1}{2}B^2
\end{multline}
while the MHD setting introduces in addition the induction equation as 
\begin{equation}
\partial_t\mathbf{B}+\nabla\cdot\left[\mathbf{v}\mathbf{B}-\mathbf{B}\mathbf{v} +\psi\mathbf{I}\right] = -\nabla\times\eta \mathbf{J} \,. \label{q-ind}
\end{equation}
For non-zero resistivity $\eta$, the resistive terms are added as ``non-local'' source terms using the compact stencil detailed in \cite{PorthXia2014}.  

To control magnetic monopole buildup numerically, we use the generalized Lagrangian multiplier (GLM) approach from \cite{Dedner2002}, where $\psi$ is an added variable that obeys
\begin{equation}
    \partial_t \psi+c_{h}^{2} \nabla \cdot \mathbf{B} = -\frac{c_h^2}{c_p^2}\psi \,.
\end{equation}
This equation has a hyperbolic advection speed $c_h$ and a parabolic factor $c_p$.
In practice, this latter equation is dealt with using the flux prescription $c^2_{\rm max}\mathbf{B}$ for the LHS, while the RHS is solved exactly over the time step $\Delta t_n$ through $\psi^{n+1}=\psi^n\exp\left(-\Delta t_n c_{\rm max}\alpha_{\rm GLM}/\min_{i=1,3}(\Delta x_i)\right)$ using second-order operator splitting. This implies availability of the maximal, global physical propagation speed $c_{\rm max}$ that also sets the time step, while $\alpha_{\rm GLM}=\min_{i=1,3}(\Delta x_i)c_h/c_p^2$ becomes a fixed parameter set to 0.5, but generally taken within [0,1] \citep{Mignone2010}.

\subsubsection{Relativistic Hydrodynamics}\label{sec:srhd}

The special relativistic hydrodynamics module solves the following set of conservation laws in flat (Minkowski) spacetime

\begin{eqnarray}
    \partial_t D + \nabla\cdot (D \mathbf{v}) = 0 \,,\label{eq:srhdD}\\
    \partial_t \mathbf{m} + \nabla \cdot (\mathbf{v}\mathbf{m} + p\mathbf{I}) = 0 \,, \label{eq:srhdm}\\
    \partial_t \tau + \nabla \cdot\left(\mathbf{m} - D \mathbf{v}\right) = 0 \,,\label{eq:srhdtau}
\end{eqnarray}
where the conserved quantities are the density $D$, the momentum-density $\mathbf{m}$ and the rest-mass subtracted total energy density $\tau:=e-D$, all evaluated in the lab-frame. The conserved variables are
\begin{equation}
    \mathbf{U} = \left(\begin{array}{cc}
        D  \\
        \mathbf{m} \\
        \tau
    \end{array}
    \right)
    = \left(\begin{array}{cc}
        \Gamma \rho \\
        \Gamma^2 \rho h \mathbf{v} \\
        \Gamma^2 \rho h - p - D
    \end{array}
    \right) \,,
\end{equation}
with $h$ being the specific enthalpy of the gas and $\Gamma$ the Lorentz factor $\Gamma=1/\sqrt{1-v^2}$.  
The primitive variables are 
\begin{equation}
    \mathbf{P} = \left(\begin{array}{cc}
        \rho  \\
        \Gamma \mathbf{v} \\
        p
    \end{array}
    \right) \,,
\end{equation}
which correspond to the fluid-frame density $\rho$, the spatial components of the four-velocity $\Gamma \mathbf{v}$ and the fluid-frame pressure $p$.  Since primitive variables are required to evaluate the fluxes, a non-linear inversion $\mathbf{P}(\mathbf{U})$ is needed.  
In \agile, we use the one introduced in \cite{BergmansKeppensEtAl2005} which expresses the energy equation as a 1D root-finding problem in $\xi = \Gamma^2 \rho h$.  
We also save the set of auxiliary variables 
\begin{equation}
    \mathbf{A} = \left(\begin{array}{cc}
        \xi  \\
        \Gamma
    \end{array}
    \right) \,,
\end{equation}
which is useful for quick conversion between $\mathbf{P}$ and $\mathbf{U}$ and to initialize an initial guess for the root-finder.  

As all Euler-type equations, the set (\ref{eq:srhdD} - \ref{eq:srhdtau}) needs to be closed with an equation of state (EOS) and we generally assume an EOS of the form $h=h(\rho,p)$.  The two options currently implemented are the \textit{ideal gas}
\begin{align}
    h(\rho,p) = 1 + \frac{\hat{\gamma}}{\hat{\gamma}-1}\frac{p}{\rho} \,,
\end{align}
and the \textit{Synge gas} of the single species trans-relativistic gas \citep{Synge1957} which is defined as 
\begin{align}
    h(\rho,p) = \frac{K_3(\theta^{-1})}{K_2(\theta^{-1})} \,,
\end{align}
with the relativistic temperature $\theta := p/\rho$ and $K_n$ being the modified Bessel function of the second kind. As in previous work \citep{MelianiSauty2004,MignonePlewa2005,RyuChattopadhyayEtAl2006,PorthOlivares2017}, we use an approximation to the previous expression that avoids the expensive evaluations of the Bessel functions \citep{Mathews1971,GoedbloedKeppens2010}.

\begin{table}
 \caption{Overview of the currently implemented physics capabilities in \agile 1.0.}
 \label{tab:physics}
 \begin{tabular*}{\columnwidth}{p{2cm}p{2cm}p{2cm}}
  \hline
  Module name & Reference & Available solvers \\
  \texttt{hd} & Section \ref{sec:hd} & \texttt{LF, HLL, HLLC} \\
  \texttt{ffhd} & Section \ref{sec:ffhd} & \texttt{LF} \\
  \texttt{mhd} & Section \ref{sec:mhd} & \texttt{LF, HLL} \\
  \texttt{srhd} & Section \ref{sec:srhd} & \texttt{LF} \\
  \hline
  \end{tabular*}
\end{table}

\section{Performance Characterization}\label{sec:performance}

\subsection{Single device benchmarks}\label{sec:singleGPU}
To establish the baseline performance on various device types, we first discuss the case of a uniform grid in standard hydrodynamics.  In particular, we set up a double-Kelvin-Helmholtz problem where the velocity switches from positive x- to negative x- and back to positive x-direction, yielding a fully periodic system.  
We adopt linear reconstruction with ``van Leer'' slope limiter, Lax-Friedrich fluxes as well as a three-step Runge-Kutta time integration.  Boundary conditions are periodic and two ghost-cells are added to each side of the blocks.  The system is advanced for 100 timesteps and we record the average number of cell updates per second (CUPS), where one ``update'' is defined as one Runge-Kutta substep.  All experiments with \agile are carried out in double precision.  

\begin{figure}
    \centering
    \includegraphics[width=\columnwidth]{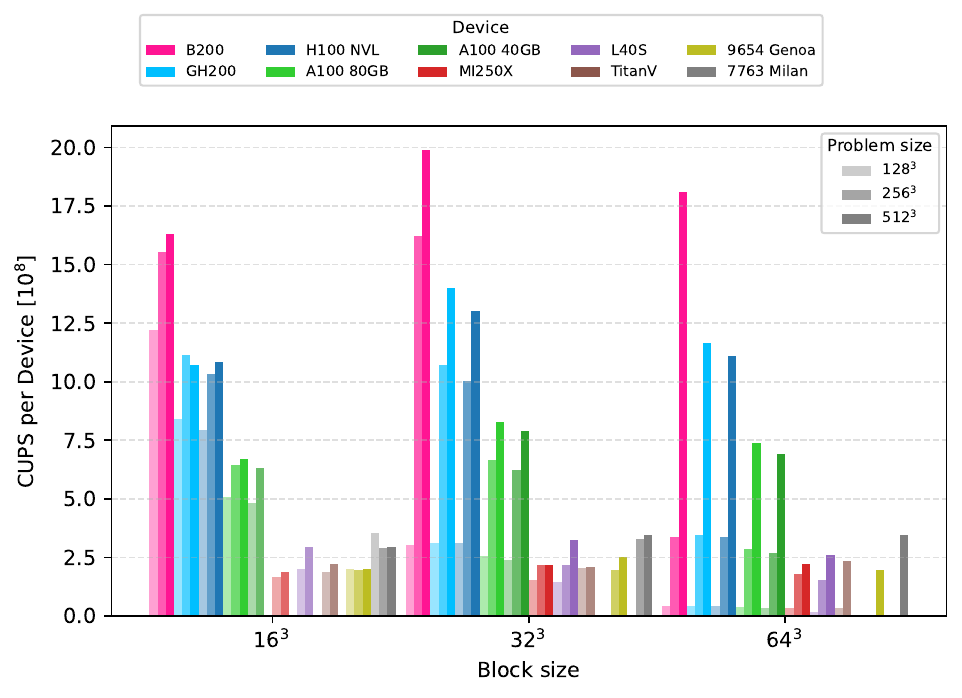}
    \caption{Single device benchmark with varying problem- and block-sizes.  Good performance is already obtained at small block size of $16^3$, a decreasing number of blocks per SM leads to reduced performance for large block sizes.  This is seen clearly for the small test case $128^3$ resulting in only 8 blocks per device when a block size of $64^3$ is used.}
    \label{fig:1GPUsummary}
\end{figure}

An overview of the performance for various devices, block- and problem-sizes is given in Figure  \ref{fig:1GPUsummary}.  
\begin{table}
 \caption{Performance on a single GPU / full CPU node, uniform grid 3D hydrodynamics with blocksize of $32^3$ and a domain size of either $256^3$ or $512^3$ cells.  See also Figure \ref{fig:1GPUsummary} and Appendix \ref{sec:singleGPUdetails} for further details, including a scan over domain and block sizes on each of these devices.}
 \label{tab:1GPUsummary}
 \begin{tabular*}{\columnwidth}{p{5cm}p{3cm}}
\hline
Device name & Performance [$10^8$ CUPS]\\
\hline
B200 "Mindwell" & 19.91 \\
GH200 "ETP" & 13.98 \\
H100 SXM5 96GB "Snellius" & 13.01 \\
A100 SXM4 80GB "wICE" & 8.28 \\
A100 SXM4 40GB "Snellius" & 7.89 \\
MI250X 1 GCD  "LUMI" & 2.16 \\
L40S & 2.11 \\
TitanV & 2.11 \\
Dual EPYC 9654 Genoa "Snellius" & 2.51 \\
Dual EPYC 7763 Milan "LUMI" & 3.41 \\
\hline
\end{tabular*}
\end{table}
Table \ref{tab:1GPUsummary} compares the main performance numbers for a fiducial block size of $32^3$ cells and a domain size of either $256^3$ or $512^3$ cells, depending on the available memory.  Our benchmark indicates that the most recent cards perform best, with the B200 reaching almost $2\times 10^9$ CUPS and the GH200 and H100 perform almost identically with $\simeq1.3\times 10^9$ CUPS, followed by the A100 with up to $8.28\times 10^8$ CUPS.  
Performance correlates strongly with the memory access speed of the devices, e.g. the G200 features 4.8TB/s, similar to the H100 SXM5 which has a nominal transfer speed of 1.97 TB/s per die (with two dies per device). This is followed by the A100 on the wICE cluster with approximately half the speed at 2.04TB/s and the A100 on Snellius with 1.55TB/s.  The AMD card on LUMI contains two Graphics Compute Dies (GCDs), each exposed as a separate GPU device. Each node therefore provides 8 GCDs across 4 MI250X cards and we use one GCD for our benchmark.  Still, the MI250X (with a transfer speed of 1.6TB/s per die) underperforms by a factor of several which might be caused by less mature support for OpenACC offloading on AMD machines \citep[see also similar findings by][]{DavisSivaramanEtAl2024}, as well as other differences in hardware such as significantly lower cache sizes (see Appendix \ref{sec:roofline} for further discussion).
It is interesting to note that the L40S which has been designed for AI applications (the vendor has not even published a double precision FLOPS value) still provides decent performance which is on par with the AMD card and the legacy TitanV. Further details on the performance of AGILE relative to the GPU peak performance are given in Appendix~\ref{sec:roofline}.
On the CPU side, AGILE also performs well on the tested multi-core EPYCs of the Genoa (192 cores per node) and Milan class (128 cores per node) with the Milan outperforming the Genoa node. 

An important consideration is the comparative cost to solution.  To give a concrete example, on the national Dutch supercomputer Snellius, the ``user exposed cost'' of one Genoa node is identical to a single H100 (ie. 192 SBU ``system billing units'' per hour).  Given the performance difference, this results in an over $5$ times reduced runtime and cost to solution when using modern H100 GPUs as opposed to the fastest CPU partition of the cluster.  For the A100 partition (128 SBU per device per hour), the cost reduction is still a worthwhile factor of $4.7$.  

To get a feeling for the minimal required load per device as well as the effect of problem granularity (block size), we repeat this test case for a range of problem sizes $\in[128^3,256^3,512^3]$ and block sizes $\in[16^3,32^3,64^3]$.  While the details are given in Appendix \ref{sec:singleGPUdetails}, the main takeaways are: a block size of $16^3$ often provides comparable and sometimes even superior performance to bigger blocks, e.g. on the H100 at $256^3$ domain, we obtain $10.32 \times 10^8$ CUPS ($16^3$ blocks) vs $10.05 \times 10^8$ CUPS ($32^3$ blocks).  This means that AGILE's update algorithm is well suited for adaptive mesh refinement simulations with moderate block sizes of $16^3$ allowing efficient local refinement.  
In terms of granularity, for a given problem size, our parameter scan indicates that the performance saturates when at least $512$ blocks per device are used. Having sufficient number of blocks (parallelized over in AGILE's ``gang loop'') often outweighs the benefits of using bigger but fewer blocks per device.

Finally, it is interesting to note that with modern large memory GPUs, the single device case already allows us to test strong scaling behavior.  Thus for the $16^3$ block size experiment, the A100-80GB (H100) indicates a strong scaling efficiency of $75\%$ ($73\%$) when the problem is scaled down from $512^3$ to $128^3$ corresponding to a factor of $64$.  Hence the under-saturated minimal load per device can be as low as $\sim 128^3$ which bodes well for multi-GPU strong scaling. 

\subsection{Multi-GPU benchmarks and scaling}

Following the single-GPUs benchmarks we investigate the performance and scaling behaviour across multiple GPUs. The same uniform grid setup as in the single-device case is used, with a fixed block size of $16^3$. We run scaling tests on the Dutch national supercomputer Snellius with four NVIDIA H100 GPUs per node and on LUMI-G with four AMD MI250X GPUs per node. For a single node, we pick the largest problem size that fits both Snellius and LUMI-G: $1024\times1024\times512$.

\begin{figure}
\centering
\includegraphics[width=\linewidth]{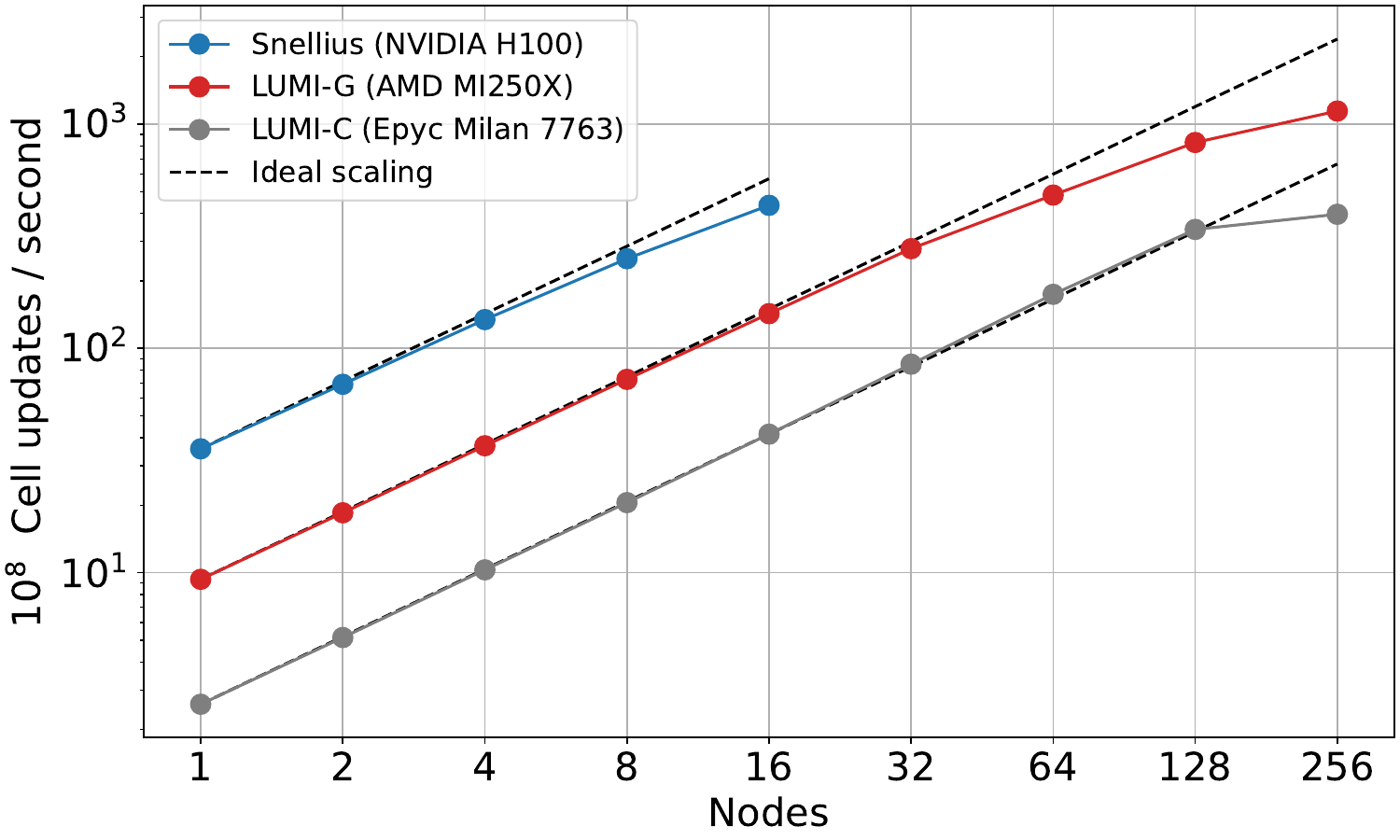}
\caption{Strong scaling. \agile scales well on both NVIDIA and AMD machines, and still does well on the LUMI CPU partition. While the absolute performance per node on Snellius is better than on LUMI, the sustained scaling and larger size of the LUMI GPU partition results in higher absolute performance.}
\label{fig:strong_scaling}
\end{figure}

\begin{figure}
\centering
\includegraphics[width=\linewidth]{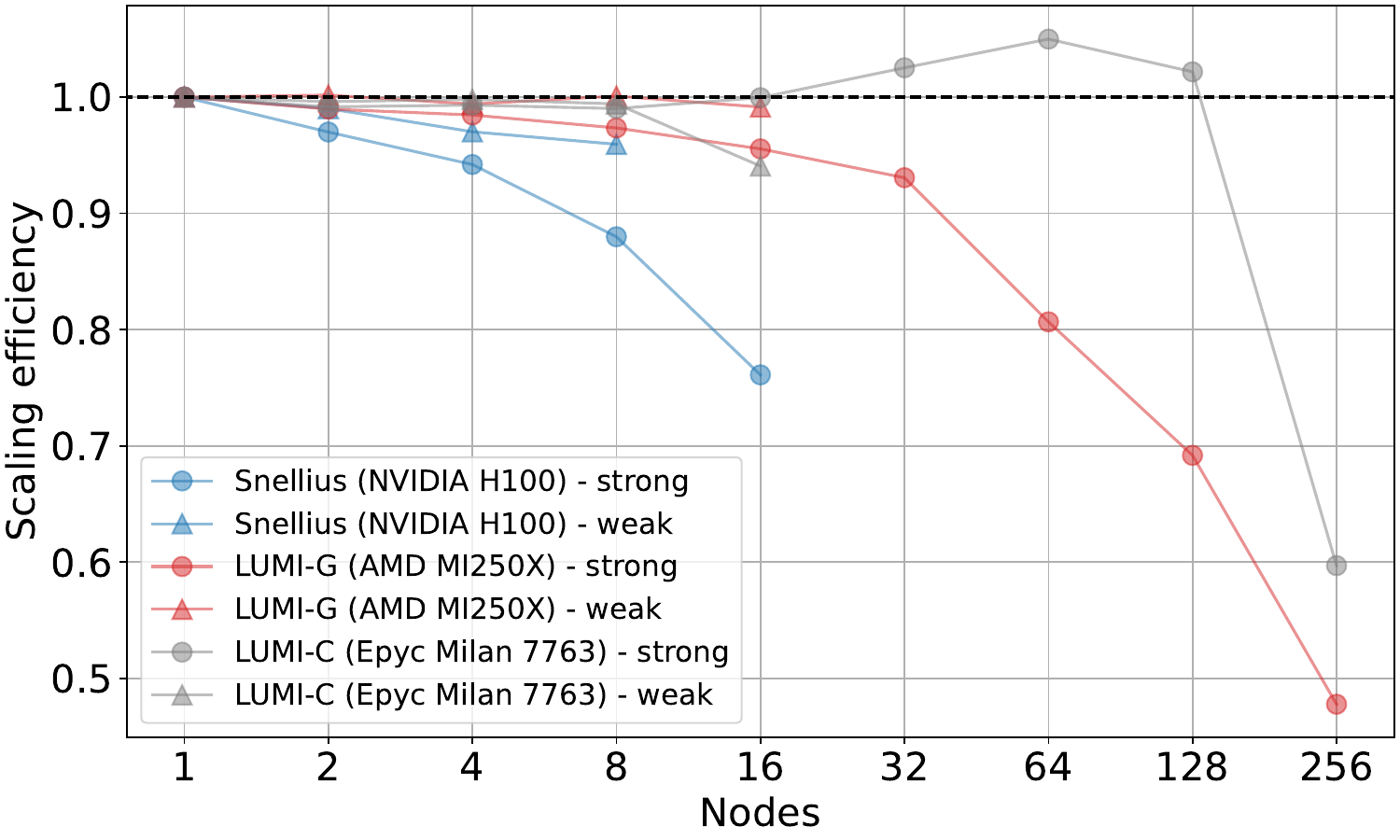}
\caption{Weak and strong scaling efficiency. \agile scales well on all systems in both strong and weak scaling.}
\label{fig:scaling_efficiency}
\end{figure}

Figure~\ref{fig:strong_scaling} shows the multi-GPU strong scaling behaviour of \agile, while in fig.~\ref{fig:scaling_efficiency} we show the scaling efficiency for both strong and weak scaling, defined as the performance relative to that of a single node. For reference, we also show the scaling behaviour on the LUMI CPU partition, where we are limited to a four times smaller problem size of 
$512\times512\times512$ on a single node. As with the single-GPU case, the H100 outperforms the MI250X and the GPU nodes are significantly faster than the CPU nodes. 
While the absolute performance for a given number of nodes is lower on LUMI-G than on Snellius, the scaling on LUMI is better, still reaching $70\%$ efficiency on 64 nodes. On Snellius we are limited to 16 nodes due to the large queuing times when requesting large portions of the cluster. On 16 nodes, the scaling efficiency is around $75\%$. On LUMI we are able to reach $10^{11}$ CUPS, using 2048 MI250X GCDs, however scaling efficiency has dropped below $50\%$ at that point. 

\section{Example applications}\label{sec:examples}

In this section we showcase some relevant applications as demonstration for the science-readiness of \agile.  We also provide quantitative measures that are compared to theoretical expectations of the non-linear evolution for the purpose of validation and cross-comparison between codes.  

\subsection{Double Mach reflection meets shock-cloud}\label{sec:wc}

Our first demonstration combines a standard test where a Mach 10 shock reflects of a wall \citep{WC1984} with a typical shock-cloud configuration encountered in many astrophysical contexts. The double Mach reflection setup was among the first to be tested with adaptive mesh refinement in 2D \citep{BC1989} and requires discretizations with shock-capturing capabilities combined with non-trivial spatio-temporally evolving boundary prescriptions. Our initial condition consists of three different, uniform states in contact:
\begin{itemize}
    \item a static ($\mathbf{v}=\mathbf{0}$) pre-shock medium, with density $\rho_{pre}=\hat{\gamma}$, $p_{pre}=1$, where our constant ratio of specific heats $\hat{\gamma}=1.4$  typical for air makes its (dimensionless) sound speed unity;
    \item a static semi-spherical region embedded in the pre-shock state with density $\rho_{c}=10$, with the sphere center positioned at $(x,y,z)=(7,0,0.5)$ and a cloud radius $r_{c}=0.25$, in pressure equilibrium, so the surface of the sphere represents a contact discontinuity;
    \item a moving post-shock state, corresponding to a Mach 10 shock. In practice, this sets the post-shock state to $\rho_{post}=8$, $p_{post}=116.5$ and post-shock velocity components $v_x,v_y$ in accord with the hydrodynamic Rankine-Hugoniot conditions, which fixes the post-shock normal velocity to $v_n=8.25$. The shock front is planar at $t=0$, and positioned according to  $x=y/\tan(\pi/3)+1/6$, so the shock makes an angle of $60^\circ$ with the horizontal $x$-direction.
\end{itemize}
We simulate this setup on a domain $[0,L_x]\times[0,L_y]\times[0,L_z]$ of size $(L_x,L_y,L_z)=(10,2,1)$. We take block sizes of $12^3$, adopt a base level grid of $240\times 48\times 24$, allowing 6 refinement levels to achieve effective resolutions of $7680\times1536\times 768$. The boundary conditions adopt the constant post-shock solution at $x=0$, extrapolate all quantities continuously at $x=10$, use reflective wall conditions at $z=0$ and $z=1$, and handle the bottom $y=0$ plane as partly fixed to the post-shock state (before $x=1/6$) and partly reflective (beyond $x=1/6$). The top boundary at $y=2$ uses the knowledge of the Mach 10 shock speed, setting the pre- and post-shock quantities in a spatio-temporally dependent fashion across the moving shock. We simulate till time $t=0.9$. Note that the original setup from \cite{WC1984} used a domain $(L_x,L_y)=(4,1)$ and followed the reflecting shock pattern till $t=0.2$. \cite{WC1984} showed that a self-similar shock reflection emerged, involving two triple points, creating a double Mach-stem reflection pattern, where a jetted flow perturbs the leading Mach stem, and Kelvin-Helmholtz roll-ups occur along a slip line and along the jet. Our cloud region is situated far enough to let it interact fully with all dynamic structures created in this double Mach reflection configuration. The cloud spherical surface is to be maintained as an exact contact discontinuity, up until the leading Mach stem arrives.

Since the only 3D effect is introduced by the semi-spherical cloud in the upstream state, the setup serves to demonstrate that we preserve exact 2D conditions up to the time where the leading Mach stem hits this region. The interaction itself shows Richtmeyer-Meshkov deformations of the shocked contact, and creates many further shock reflections throughout. A turbulent region develops that eventually passes across the rightmost open boundary without any obvious artificial reflections. We also maintain symmetry about the vertical midplane $z=0.5$ for a very long time into this turbulent interaction.

\begin{figure}
\centering
\includegraphics[width=\linewidth]{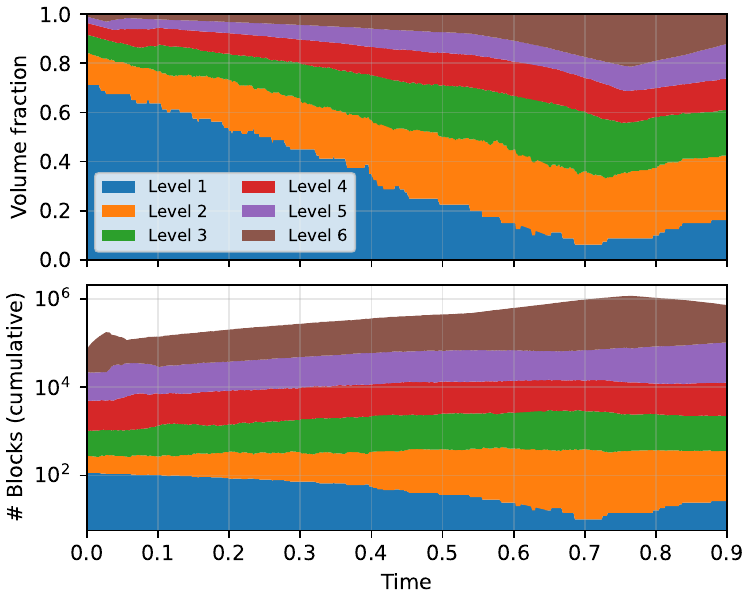}
\caption{Evolving AMR level population over simulation time for the double Mach reflection simulation.  As more of the initially uniform domain becomes processed by the shock, the level 1 occupancy shrinks in favor of refined grids.  At peak, over a million grid blocks are activated which decreases again as the shocked cloud is pushed out of the computational domain.}
\label{fig:WClevelpop}
\end{figure}

The simulation uses the Euler equations without any source terms, discretized in a three-step integrator using HLLC and Van Leer slope limiting. The CFL parameter was set to 0.8 and automatic refinement is activated based on density only.  
Since the simulation increases dramatically in memory footprint during runtime, starting at snapshot sizes of 4.7 GB and peaking at 76 GB, the simulation was restarted several times with progressively increased node count (ranging from 64 nodes to 256 nodes -- resulting in up to 2048 GPUs).  The progression of the AMR level population is further shown in Figure \ref{fig:WClevelpop}; the $t=0$ frame has 116 base level blocks versus 50304 level 6 blocks, while the final $t=0.9$ frame counts 26 base level blocks and 631680 level 6 blocks.  At peak, over one million AMR blocks are handled, resulting in over $1.7\times 10^9$ computational cells.

\begin{figure*}
\centering
\includegraphics[width=1\linewidth,trim={2.5cm 0.5cm 1cm 0},clip]{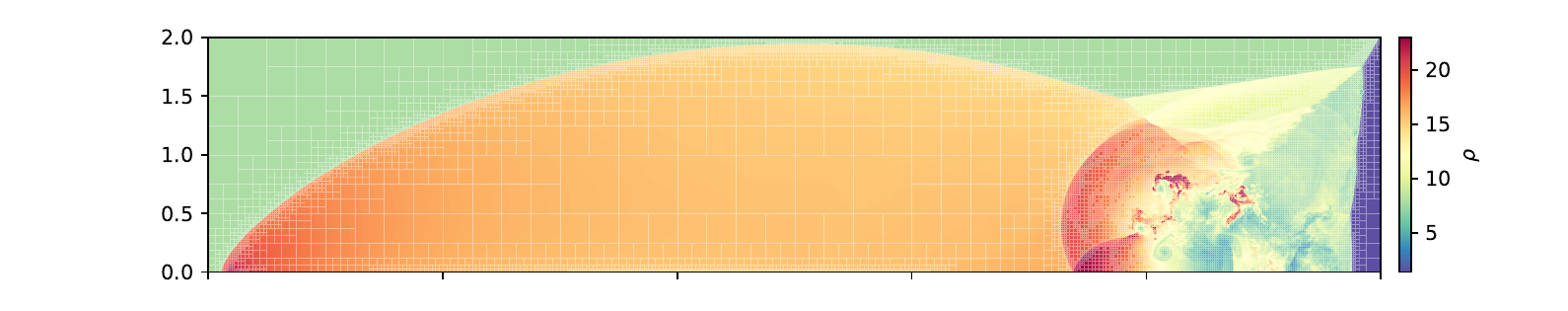}
\includegraphics[width=1\linewidth,trim={2.5cm 0 1cm 0.5cm},clip]{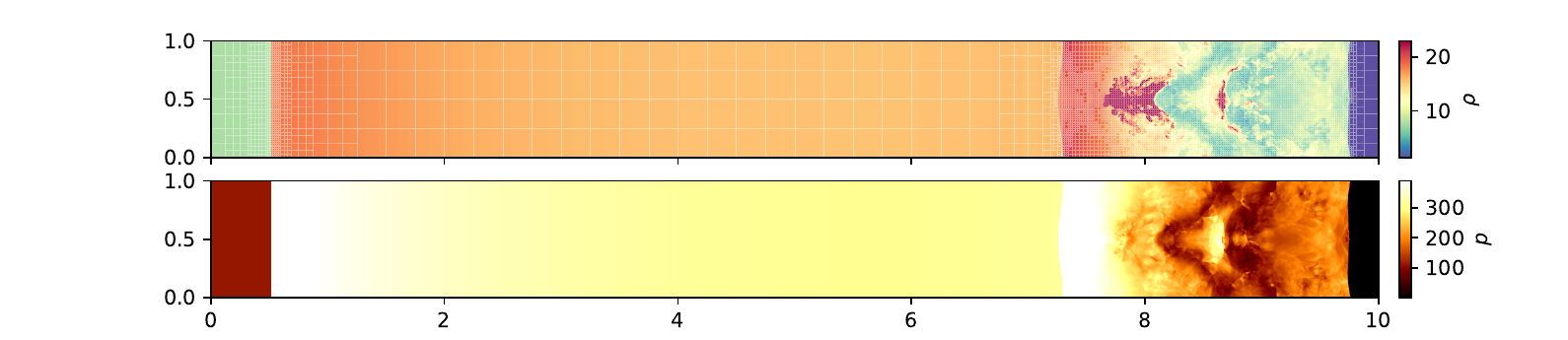}
\caption{Double Mach reflection structure impacting on a dense cloud.  This case shows dynamical adaptive mesh structures using a total of six refinement levels.  The simulation used up to 2048 GPUs (256 nodes) on the LUMI cluster.  
See \url{https://surfdrive.surf.nl/s/eYjk7fzaZnwXX8X} for an animated version of the figure. We show the density at $t=0.75$ in a vertical slice at $z=0.25$ (top) and density (middle) and pressure (bottom) in the horizontal slice $y=0.5$.}
\label{fig:DoubleMach}
\end{figure*}

During runtime, we produce log-file output at high cadence ($\Delta t=0.001$) collecting cumulative information such as the volume-integrated total mass and energy. Next to native \texttt{.dat} snapshot files that contain all conservative variables on the hierarchical AMR tree, which allow full restarts, we obtain 3 data slices in \texttt{.vtu} format on three selected cutting planes at an intermediate cadence of $\Delta t=0.005$. These sequences can directly be visualized in any modern visualization tool, such as \textit{ParaView}, \textit{VisIt} or \textit{yt}. An impression of the evolution is shown in Fig \ref{fig:DoubleMach}. The figure shows the density in both a vertical slice taken at $z=0.25$ (top panel) and a horizontal slice at $y=0.5$ (middle). The bottom shows the pressure distribution in the same $y=0.5$ slice. An animation of the evolution is also provided.

Note that the slice data can directly store primitive variables, augmented with any number of user-specific derived variables: for the case shown here, we augmented all slice output with a Schlieren variation of the density gradient.  
The entire run, involving four consecutive batch-scripted jobs completed within 22 hours and 10 minutes wall-clock time on LUMI, of which 11 hours on 64 nodes, 4.5 hours on 128 nodes, and the remainder on 256 (totaling in 2944 node hours). We note in passing that restarts can even be done on different architectures.

\subsection{Multiphase turbulent mixing}

As a first science application, we here consider the turbulent mixing of a multi-phase gas. Turbulent (radiative) mixing layers are expected wherever cool, dense gas is in shear contact with a much hotter, more tenuous phase. In the interstellar medium (ISM) this occurs, for example, at interfaces between supernova-heated coronal gas and warm or cold clouds \citep{1990MNRAS.244P..26B, SlavinShullEtAl1993}, and in the interaction of high-velocity clouds with the Galactic corona \citep{SembachWakkerEtAl2003}. Similar conditions are ubiquitous in the circumgalactic and intergalactic medium, where cool clouds and filaments are embedded in hot virialized halos and galactic winds \citep[e.g.][]{FabianSandersEtAl2003,KwakShelton2010,JiOhEtAl2019, gronke_growth_2018,Fielding_2020,LochhaasBryanEtAl2020}.  

Turbulent mixing of two adiabatic fluids with temperatures $T_\mathrm c$ and $T_\mathrm h$ is expected to produce an intermediary mixing phase of temperature $T_*=\sqrt{T_\mathrm hT_\mathrm c}$, through the Kelvin-Helmholtz instability \citep{1990MNRAS.244P..26B,SlavinShullEtAl1993}. If furthermore this intermediate phase experiences relatively strong optically thin radiative cooling, due to a cooling curve $n^2\Lambda(T)$ peaked at $T_*$, the intermediate phase becomes thin and fractal in nature, with the fractal dimension conjectured to approach $5/2$ \citep{Fielding_2020}.
The energy lost to cooling is balanced by an enthalpy transfer from the hot to the cold phase through the so-called `turbulent radiative mixing layer' (`TRML') and the characterization of this fractal structure is essential to model interstellar cooling processes \citet[][]{Fielding_2020}.

This 3D hydrodynamic setup illustrates the user-adaptable physics of \agile with a custom radiative cooling curve and boundary conditions, as well as fine-grained control over numerical aspects such as user-defined static mesh refinement.
We use the HLLC solver, Van Leer slope limiter and a three-step Runge-Kutta time-update. The effective resolution is $768\times768\times6144$ realized by 2 grid levels, counting 2944 level 1 and 9216 level 2 blocks of sizes $48^3$.
A high resolution region in the center is specified, in which the TRML initially forms near the bottom, and gradually migrates upwards as enthalpy is transferred.
The numerics are similar to Section \ref{sec:cloud_crushing}: MUSCL reconstruction with a Van Leer slope limiter, the HLLC flux scheme, and threestep Runge-Kutta time integration.

\begin{figure}
    \centering
    \includegraphics[width=\linewidth]{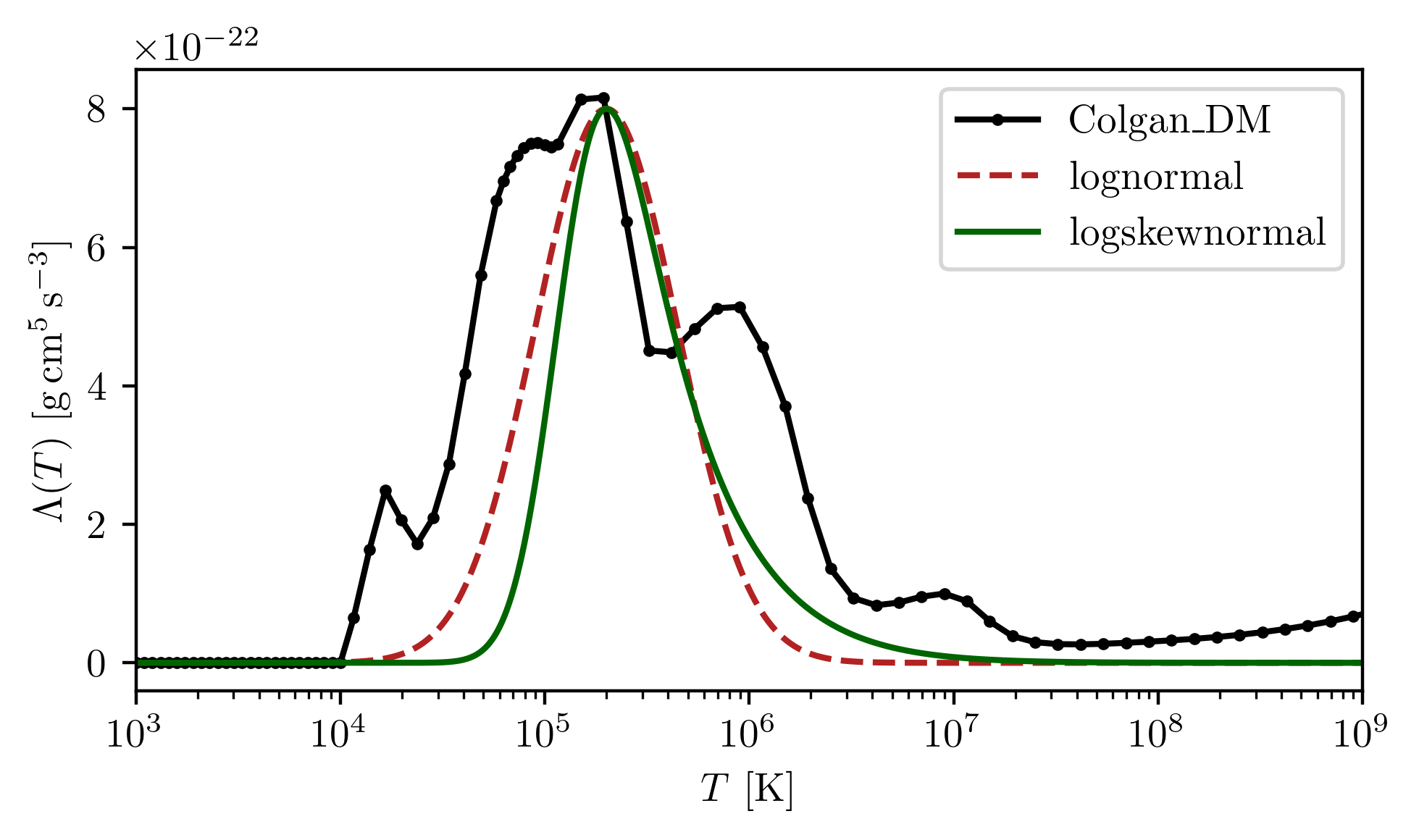}
    \caption{A log-skew-normal cooling curve acts as a generic, parameterizable alternative to the Colgan-DM curve. Skew is necessary to tame the cooling in the dense phase (as it does for Colgan-DM), producing a sufficiently symmetric peak in $n^2\Lambda(T)$. This particular log-skew-normal curve is the one used in the setup.}
    \label{fig:cooling_curves}
\end{figure}

The Kelvin-Helmholtz setup is based on the standard 2D noisy benchmark in \cite{Lecoanet_2015} appropriately generalized to 3D, in particular with a root mode $(\sin(2\pi x/L)+\sin(2\pi y/L))/2$, and additional 2D band-limited uncorrelated white noise around the interface, as opposed to 1D noise. The TRML setup is based on \cite{Fielding_2020}, with a hand-crafted cooling curve to elucidate the effect of localized cooling.
In our case, $\Lambda(\ln T)$ is taken to be a skew-normal distribution, which well approximates the right skew of cooling curves such as \amrvac's Colgan-DM; see Figure \ref{fig:cooling_curves}.
The curve is chosen such that 
\begin{align*}
    n_\mathrm c^2\Lambda(T_\mathrm c)\approx n_\mathrm h^2\Lambda(T_\mathrm h),
\end{align*}
i.e. the initial dense cold phase radiates similarly to the hot rarefied phase, with some peak in-between.
The curve is based on the normalized log-skew-normal distribution with density
\begin{align*}
    \Lambda(T)=\frac{2 A}{\sigma (T/\mathrm K)}\phi\left(\frac{\ln(T/\mathrm K)-\mu}\sigma\right)\Phi\left(\alpha\frac{\ln(T/\mathrm K)-\mu)}\sigma\right),
\end{align*}
with $\alpha$ the skew parameter, $\mu$ and $\sigma$ the familiar Gaussian parameters for $\alpha=0$ and $A$ a normalization factor.
Here $\Phi$ is the Gaussian distribution and $\phi(x)=d\Phi(x)/dx$.
Having astrophysical applications in mind, we choose a mixing temperature $T_*=0.2\,\mathrm{MK}$, cold phase density $\rho_\mathrm c=10^{-20}\,\mathrm{g\,cm^{-3}}$, cooling peak
$\max\Lambda=\Lambda(T_*)=8\times10^{-22}\,\mathrm{g\,cm^5\,s^{-3}}$ at $\text{argmax}_T\,\Lambda(T)=T_*$, normal standard deviation $\sigma=3$ (in $\ln T$-space), and skew parameter $\alpha=7$. By fitting, this corresponds to $\mu\approx1.19$ and $A\approx8.19\times10^{-16}\,\rm g\, cm^5\, s^{-3}$.
This curve is shown in Figure \ref{fig:cooling_curves}.

As in \cite{Fielding_2020}, the density/temperature contrast between the hot and cold phases is $\chi=100$, the relative dominance of optically thin radiative cooling to shear flow (in time scales) is $\xi=10$ in the mixing layer, and the hot phase Mach number is $0.3$, while the cold phase is supersonic.
The characteristic shear flow length scale and time scale are implied to be $2.03\times10^{16}\,\mathrm{cm}$ and $2.88\times10^9\,\rm s$ respectively.
In the initial condition, $n^2\Lambda(T)$ aggressively increases by 3 orders of magnitude around the mixing layer.

\begin{figure*}
    \centering
    \includegraphics[width=.9\linewidth]{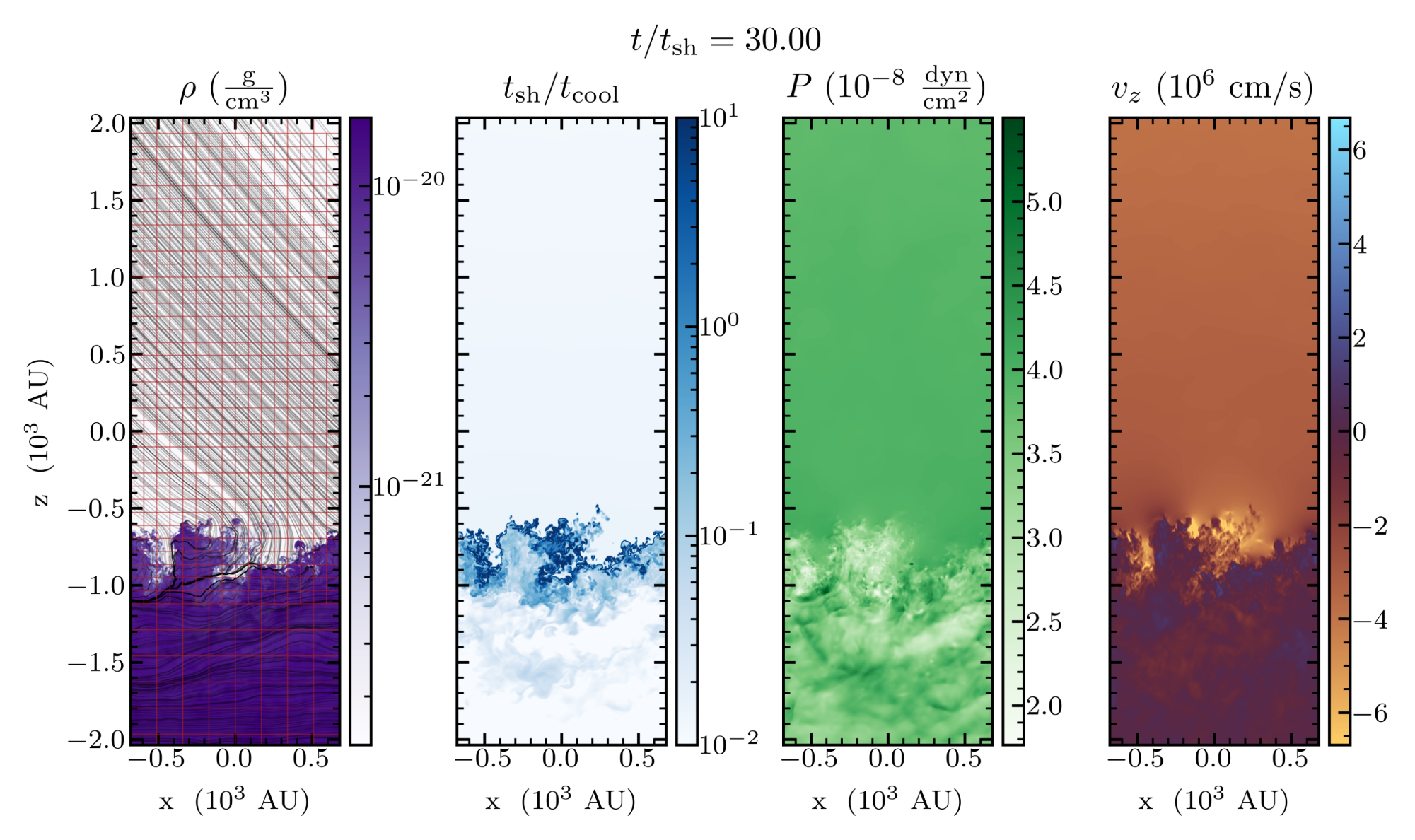}
    \caption{A slice of the turbulent shearing system showing the TRML. The absorption of material from the hot phase is characterized by a steady diagonal velocity field. The $\rho$ slice shows SMR blocks boundaries in red, the TRML has room to migrate upwards.  See \url{https://surfdrive.surf.nl/s/6S9wrHE9E8Nj3rn} for an animated version of the figure. }
    \label{fig:TRML_slices}
\end{figure*}

\begin{figure}
    \centering
    \includegraphics[width=\linewidth]{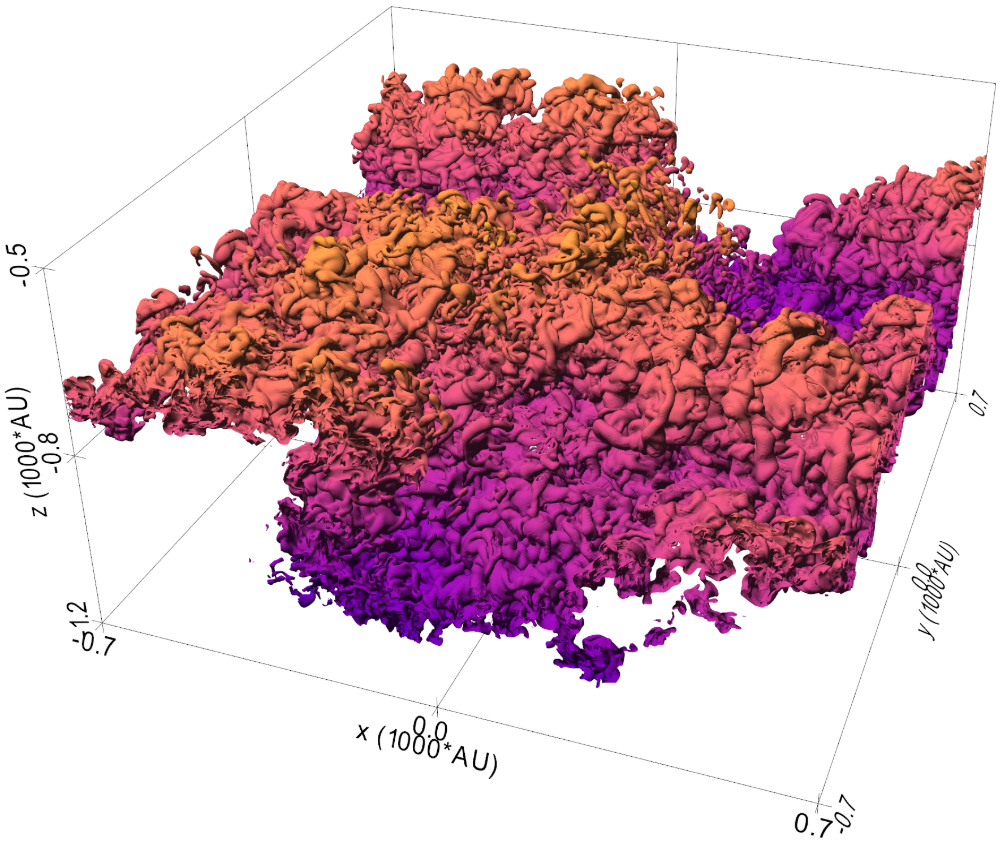}
    \caption{A $t_\mathrm{sh}/t_\mathrm{cool}=5$ isosurface, containing the idealized $t_\mathrm{sh}/t_\mathrm{cool}=\xi$ TRML, and displaying its fractal nature. The color reflects the vertical coordinate. Based on Figure 3 in \protect\cite{Fielding_2020}.
    }
    \label{fig:TRML_iso}
\end{figure}

\begin{figure}
    \centering
    \includegraphics[width=\linewidth]{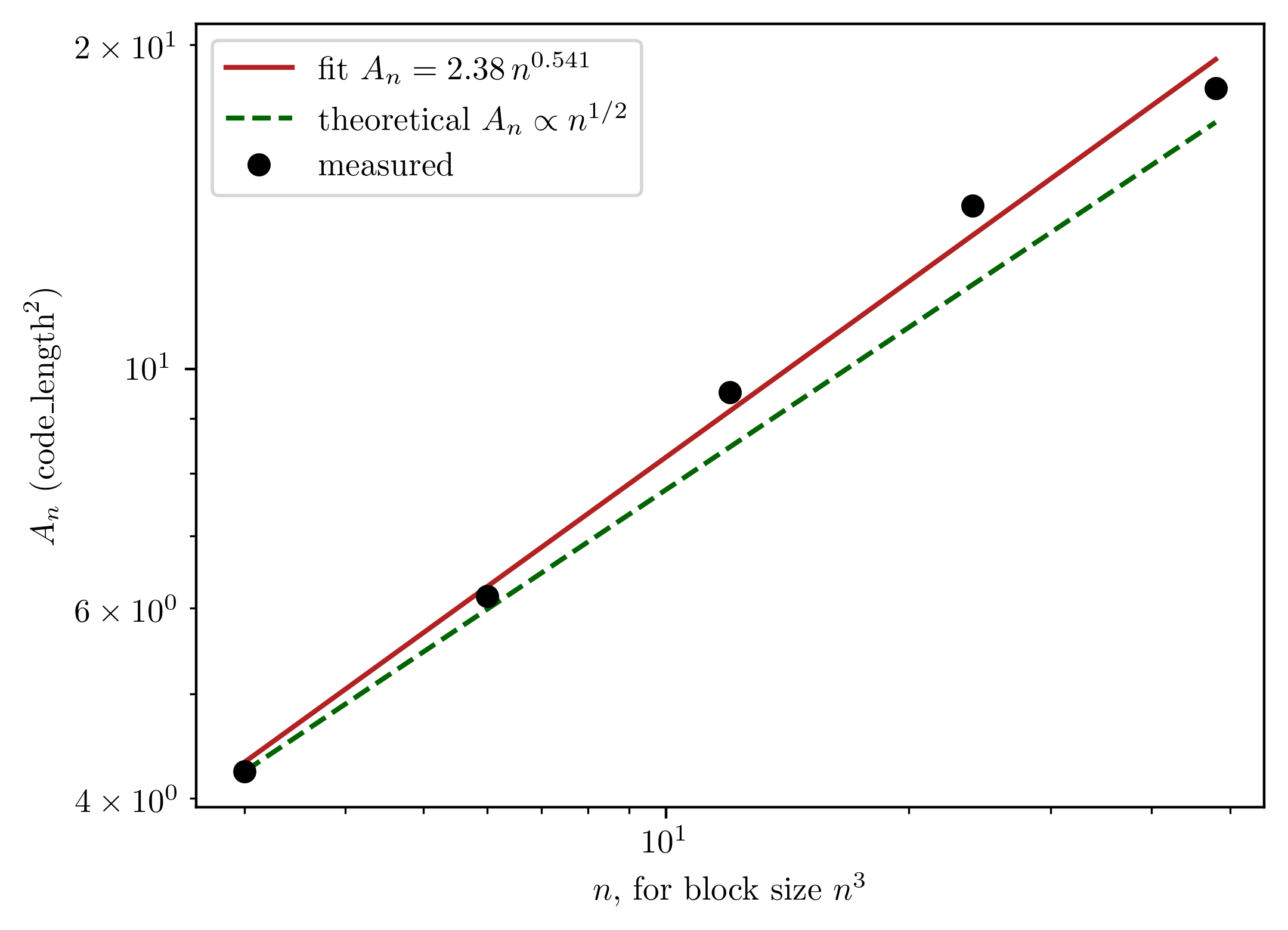}
    \caption{Measured fractal dimension $d=2.541$, for the same time $t/t_\text{sh}=30$ shown in Figures \ref{fig:TRML_slices} and \ref{fig:TRML_iso}
    }
    \label{fig:TRML_fit}
\end{figure}

Figure \ref{fig:TRML_slices} shows the evolution of the system after 30 shearing times. 
A thin mixing layer has developed which pulls material from the hot phase. Material from the hot phase flows into the mixing layer which itself moves upwards as more material is mixed into the cold phase.  A comparison of the cooling and shear timescales demonstrates that cooling is indeed dominated by the thin mixing layer which shows a fractal structure.  To further quantify the mixing surface, we here compute its fractal dimension.  The size $A_\ell$ of a (fractal) surface depends on the measuring scale $\ell$ as $A_\ell\propto \ell^d$, with $d$ its fractal dimension. By defining the TRML as a region with a relatively high ratio of cooling strength to shear flow strength ($t_\mathrm{sh}/t_\mathrm{cool}$, see second panel of Figure \ref{fig:TRML_slices}), we compute its fractal dimension $d$ by subsampling the AMR blocks and counting cells that meet the criterion, or alternatively computing the area of isosurfaces through a $t_\mathrm{sh}/t_\mathrm{cool}$ threshold. Either method is consistent with the dimensionality $d=5/2$, if the TRML has sufficiently equilibrated.  Figure \ref{fig:TRML_iso} shows a rendering of such an isosurface while Figure \ref{fig:TRML_fit} shows the extraction of the scaling with $d=2.541$.  

In terms of performance, the simulation discussed here ran on 16 LUMI MI250X nodes with $\gtrapprox1.5\times10^8$ CUPS per GCD, close to the uniform-grid single-GPU benchmark discussed in Section \ref{sec:singleGPU}.

\subsection{Cloud crushing} \label{sec:cloud_crushing}
As a demonstration of \agile's hydrodynamics module with radiative cooling and adaptive mesh refinement, we simulate the interaction of a cold dense gas cloud with a hot wind --- a process known as \textit{cloud-crushing} \citep{klein_hydrodynamic_1994}.
In this scenario, radiative cooling of the mixed gas can counteract the hydrodynamic disruption of the cloud caused by the inflowing wind \citep{gronke_growth_2018}.
We follow the setup of \citet{girichidis_situ_2021}, adopting a 3D rectangular Cartesian box with continually inflowing wind from the left boundary of temperature $T_w = 10^6~\rm{K}$, number density $n_w = 10^{-3}~\rm{cm}^{-3}$, and velocity $v_w = 100~\rm{km}~\rm{s}^{-1}$.
The cloud has radius $r_c = 50~\rm{pc}$ and is initialised in thermal pressure equilibrium with the ambient medium at three density contrasts $\chi \equiv \rho_c/\rho_w \in \{140, 240, 340\}$; the resulting cloud temperatures, densities, and characteristic timescales are summarised in Table~\ref{tab:cc-params}.
The wind drives a shock into the cloud interior at Mach number $\mathcal{M}_c = v_w/c_{s,c} \approx 10$, compressing it while simultaneously triggering Kelvin--Helmholtz and Rayleigh--Taylor instabilities at the cloud-wind interface that act to ablate and mix the cloud material into the wind.
The \textit{cloud-crushing time} \citep{klein_hydrodynamic_1994} 
\begin{equation}
    t_\mathrm{cc} = \frac{\sqrt{\chi}\,r_c}{v_w},
    \label{eq:tcc}
\end{equation}
sets the characteristic timescale for cloud compression and destruction, while the \textit{gas cooling time} \citep{girichidis_situ_2021}
\begin{equation}
    t_\mathrm{cool,c} \sim \frac{k_BT_c}{n_c\Lambda(T_c)},
    \label{eq:tcool}
\end{equation}
sets the timescale for radiative losses; the ratio $t_\mathrm{cool,c}/t_\mathrm{cc}$ provides a useful indication of the importance of cooling.
A turbulent velocity field with a Burgers power spectrum ($P(k) \propto k^{-4}$, $v_\mathrm{rms} = 1~\rm{km~s}^{-1}$) and a small density asymmetry are seeded inside the cloud to break symmetry and promote realistic mixing.
The problem domain spans $20r_c \times 4r_c \times 4r_c$ with the cloud centred at $x = 5r_c$.
Radiative cooling is prescribed by the \texttt{Colgan\_DM} curve --- a composite tabulation consisting of the curves of \citet{colgan_radiative_2008} at high temperatures and \citet{dalgarno_heating_1972} at low temperatures.

\begin{table}
    \centering
    \caption{Cloud and wind parameters for the three density contrast runs.
    The cloud temperature $T_c = T_w/\chi$ and number density $n_c = \chi n_w$ follow from pressure equilibrium with the wind ($T_w = 10^6~\rm{K}$, $n_w = 10^{-3}~\rm{cm}^{-3}$). Timescales are defined in Equations~\ref{eq:tcc} and \ref{eq:tcool}.}
    \label{tab:cc-params}
    \begin{tabular}{ccccccc}
        \hline
        $\chi$ & $T_c$ (K) & $n_c$ (cm$^{-3}$) & 
        $t_\mathrm{cc}$ (Myr) & $t_\mathrm{cool,c}$ (Myr) & 
        $t_\mathrm{cool,c}/t_\mathrm{cc}$ \\
        \hline
        140 & 7143 & 0.14 & 6.2 & 5.6 & 0.90 \\
        240 & 4167 & 0.24 & 7.6 & 3.8 & 0.50 \\
        340 & 2941 & 0.34 & 9.2 & 2.0 & 0.22 \\
        \hline
    \end{tabular}
\end{table}

\begin{figure*}
    \centering
    \includegraphics[width=1\linewidth]{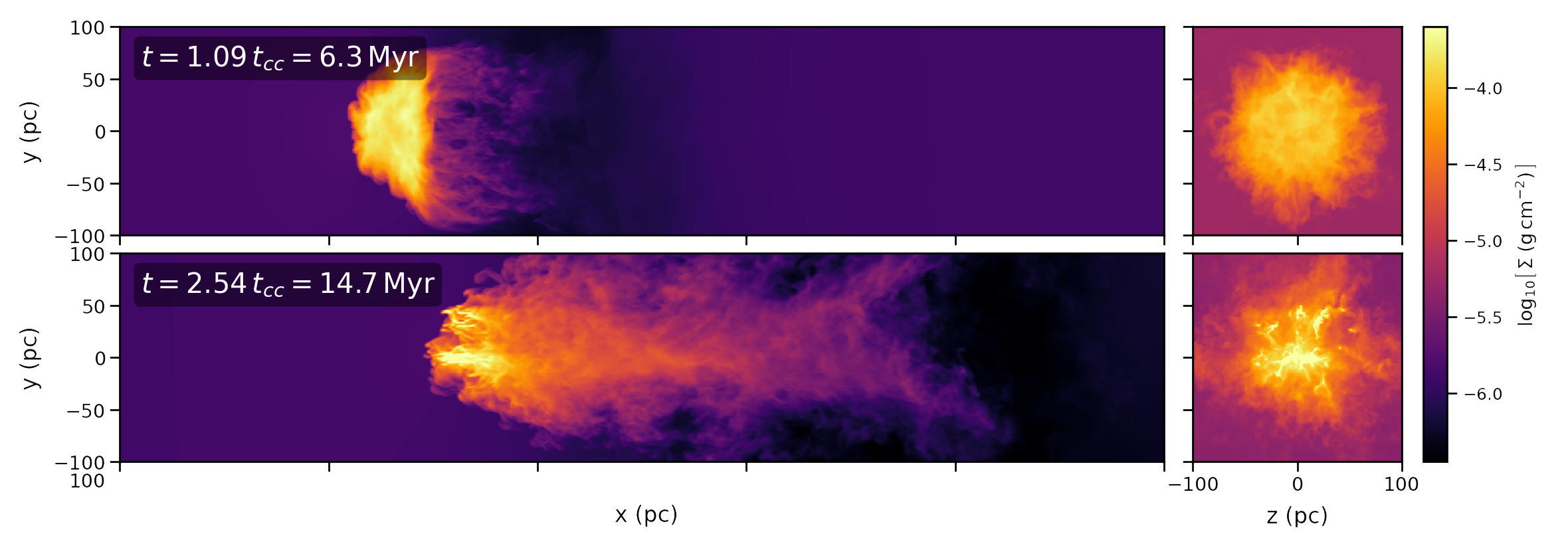}
    \caption{Cloud morphology and AMR grid structure for the $\chi = 240$ cloud-crushing run.
    Projected column density $\log_{10}\Sigma$ along the $z$-axis (left panels) and face-on $x$-axis projection (right insets) at $t \simeq 1\,t_\mathrm{cc} = 6.3~\rm{Myr}$ (top row) and $t \simeq 2.5\,t_\mathrm{cc} = 14.7~\rm{Myr}$ (bottom row). 
    }
    \label{fig:cc-cloud_morphology}
\end{figure*}

\begin{figure*}
    \centering
    \includegraphics[width=1\linewidth]{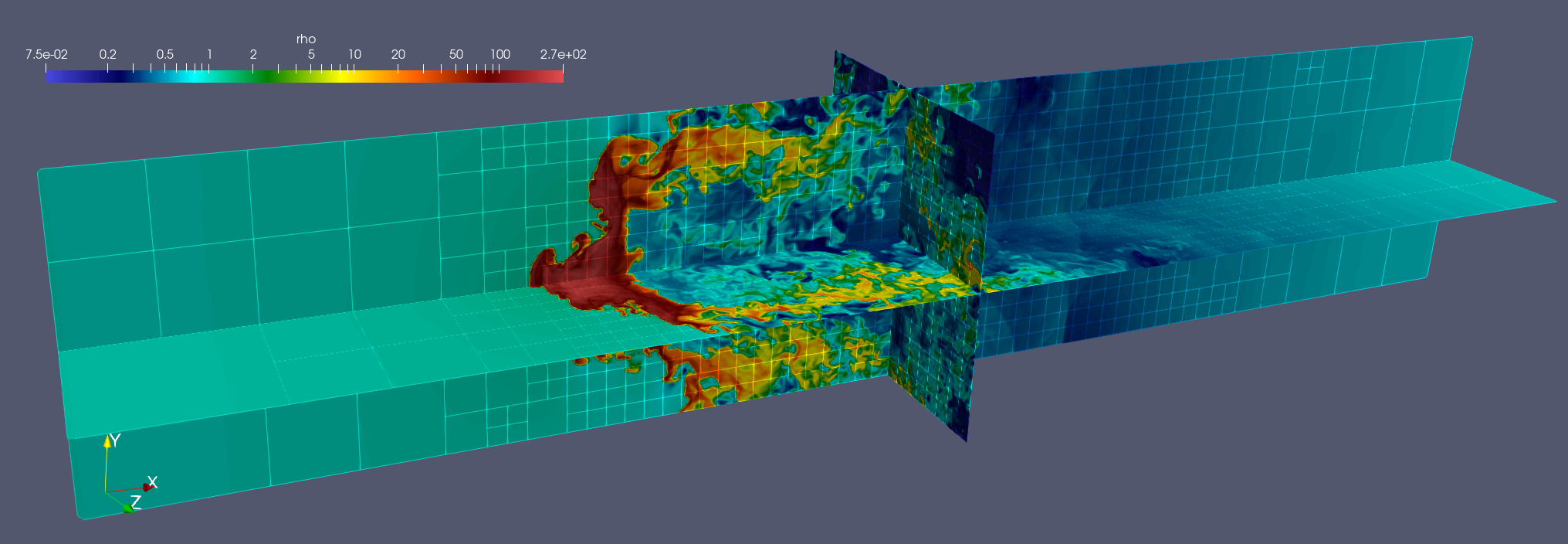}
    \caption{Density slices through the $\chi=240$ cloud-crushing simulation at $t \simeq 1.86\,t_\mathrm{cc} = 10.7~\rm{Myr}$, computed with four AMR levels and the TVDLF scheme, showing the shock-compressed cloud core (red) and the turbulent, KH/RT-disrupted tail mixing into the ambient wind (teal). See \url{https://surfdrive.surf.nl/s/5KGAx4Ct8E7anwE} for an animated version of the figure. 
    }
    \label{fig:cc-cloud_3d}
\end{figure*}

Figure~\ref{fig:cc-cloud_morphology} shows the column density (side-on and face-on) for the $\chi = 240$ run, performed with linear (MUSCL) reconstruction using a Van Leer slope limiter \citep{van_leer_towards_1974}, HLLC Riemann solver, and three-stage Runge--Kutta time integration, on two NVIDIA L40S GPUs.
The computational domain is discretised on a base grid of $320 \times 64 \times 64$ cells with three AMR levels (block size $32^3$, up to 4096 blocks), yielding an effective resolution of $1280 \times 256 \times 256$.
Mesh refinement follows L\"{o}hner's scheme \citep{lohner_1987} applied to a weighted combination of the passive tracer quantity (80\%) and density (20\%), with additional coarsening enforced at the inflow boundary.
At $t \simeq 1~t_\mathrm{cc}$ the cloud has been shock-compressed, forming a dense core, while hydrodynamic instabilities at the cloud-wind interface are actively ablating the cloud surface with stripped material forming a downstream tail.
By $t \simeq 2.5\,t_\mathrm{cc}$ the tail has extended across a large portion of the full domain length and the cloud has undergone significant mass stripping, with the remaining dense core also accelerated in the direction of the wind flow.
Radiative cooling in the mixed layer at the cloud-wind interface acts to inhibit the hydrodynamic disruption that would otherwise destroy the cloud within a few $t_\mathrm{cc}$ \citep{gronke_growth_2018, girichidis_situ_2021}.
The face-on projections (right panels) confirm the genuinely three-dimensional, asymmetric morphology seeded by the turbulent initial conditions.

Figure~\ref{fig:cc-cloud_3d} shows three orthogonal density slices through the same $\chi=240$ setup at $t \simeq 1.86\,t_\mathrm{cc}$, run at double the effective resolution ($2560\times1024\times1024$, four AMR levels) using the TVDLF scheme. 
The AMR grid structure is visible directly in the slices, with refinement automatically concentrated on the cloud and its turbulent wake.

\begin{figure}
    \centering
    \includegraphics[width=1\linewidth]{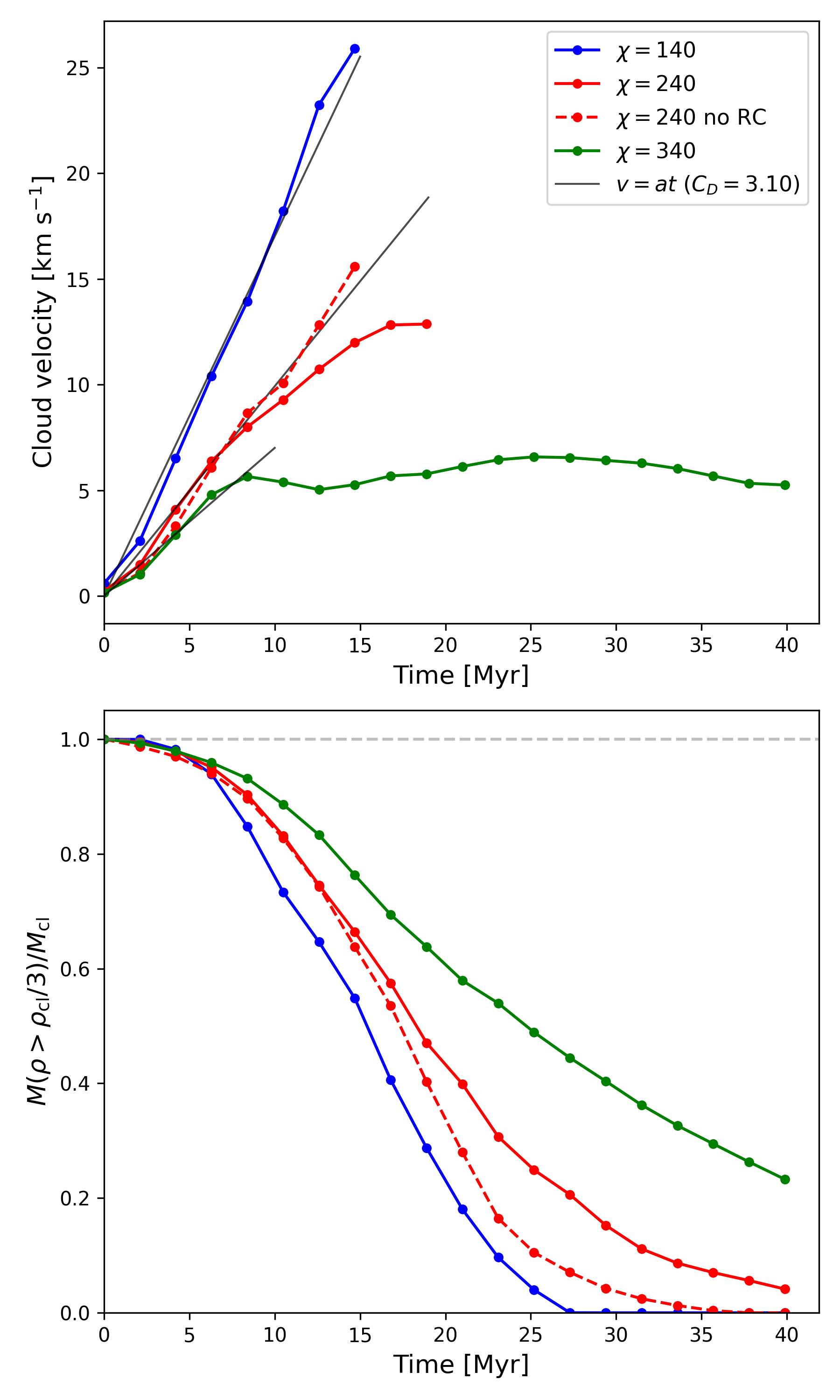}
    \caption{Quantitative diagnostics for cloud-crushing runs at $\chi \in \{140, 240, 340\}$ on a uniform grid of $1280 \times 256 \times 256$ cells using the HLL solver, with an additional no-radiative-cooling run at $\chi = 240$ (dashed red).
    \textit{Top}: cloud velocity $v_{25}(t)$, defined as the velocity at the 25th percentile of the 
    integrated mass distribution along $x$; gray lines show the analytic ram-pressure drag prediction.
    \textit{Bottom}: cold cloud mass fraction $f_\mathrm{cl}(t)$.}
    \label{fig:cc-quantitative}
\end{figure}

Figure~\ref{fig:cc-quantitative} shows quantitative diagnostics for all three density contrasts, drawn from uniform-grid runs at resolution $1280 \times 256 \times 256$ on 32 MI250X GCDs on LUMI 
, this time using a HLL Riemann solver.
We additionally include a run without radiative cooling ($\chi = 240$, dashed red) to isolate the effect of the cooling process.
We measure the cloud velocity as $v_{25}(t)$, the velocity at the 25th percentile of the integrated mass distribution along $x$ \citep{girichidis_situ_2021}.
Curves are shown only while the slope remains positive, as the percentile indicator becomes unreliable once the cloud is sufficiently disrupted.
The gray lines show the analytic ram-pressure drag prediction $v_\mathrm{cl}(t) = at$ with constant acceleration $a = 3C_D v_w^2 / (8\chi r_c)$, derived under the assumption $v_\mathrm{cl} \ll v_w$, with drag coefficient $C_D$ fitted to the early linear regime of each cooling run.
The fitted drag coefficients are consistent across all three density contrasts, $C_D \in \{3.06, 3.06, 3.19\}$ respectively, confirming that the early acceleration follows the expected $1/\chi$ scaling of the ram-pressure drag law across the full parameter range.
\citet{girichidis_situ_2021} adopt $C_D = 1$ for their analytic estimates and find that their simulated accelerations fall below this by a factor of a few, attributed to fragmentation reducing the effective cross-section; our higher value $C_D \approx 3.1$ also reflects the deformation and ablation of the cloud increasing the effective cross-section above $\pi r_c^2$ in the absence of stabilizing magnetic fields and self-gravity.
The \textit{cold cloud mass fraction} \citep{gronke_growth_2018}
\begin{equation}
    f_\mathrm{cl}(t) = \frac{M(\rho > \rho_\mathrm{cl}/3)}{M_\mathrm{cl}} 
\end{equation}
tracks the fraction of the original cloud material remaining above a density threshold of $\rho_\mathrm{cl}/3$, capturing gas that has been either advected out of the domain or mixed below the cold-phase threshold by hydrodynamic instabilities.
This decreases monotonically for all runs, indicating that radiative cooling slows but does not completely prevent cloud disruption in the chosen $t_\mathrm{cool,c}/t_\mathrm{cc} < 1$ parameter range.
The comparison between the $\chi = 240$ runs with and without radiative cooling directly demonstrates the effect of the cooling process: the cloud without cooling is accelerated faster and more quickly disrupted, confirming that \agile's hydrodynamics and cooling modules reproduce the expected impact of radiative cooling on cloud evolution.

We recorded the performance and wall-clock time for these simulations. 
The uniform grid LUMI runs ($1280\times256\times256$ cells, HLL solver, 32 MI250X GCDs) achieved between $1.02$–$1.22\times10^{8}$ cell updates per second per GCD across all four runs (three cooling cases plus the no-cooling comparison), averaging $1.12\times10^{8}$ cell updates/s/GCD, corresponding to an aggregate throughput of $\sim 3.6\times10^{9}$ cells/s across the full run.
The corresponding average wall-clock time was approximately 59 minutes.
The 3-level AMR run on two NVIDIA L40S GPUs (HLLC solver) achieved $1.38\times10^{8}$ cell updates/s per GPU, corresponding to a total throughput of $2.76\times10^{8}$ cells/s, at a wall-clock time of approximately 3 hours and 49 minutes.

\subsection{Multiphase dynamics in the solar corona}\label{sec:rain}

As a demonstration of \agile's three-dimensional FFHD capability with thermodynamic source terms, we consider multiphase evolutions in a structured coronal magnetic field. As recently reviewed in \cite{KeppensZhouXia2025}, the optically thin radiative loss channel in the tenuous and hot ($>1$ MK) coronal plasma can trigger runaway condensation formations, including prominences \citep{XiaChenKeppens2012,KeppensXia2014,Zhou2024,Zhou2025} and coronal rain \citep{JercicJenkinsKeppens2024}. Prominences as well as coronal rain features are cool, dense condensations suspended in the hot corona, so their formation provides a stringent test of whether a numerical framework can simultaneously capture chromospheric evaporation and thermal instability development, involving radiative cooling, gravity, anisotropic thermal conduction, and magnetic support.

We follow the evaporation-condensation prominence models developed by \citet{XiaChenKeppens2012} and \citet{KeppensXia2014}, while also connecting to the later studies of \citet{JercicKeppens2023} and \citet{JercicJenkinsKeppens2024}. In the present setup, both the evaporation-condensation scenario and the in-plane magnetic topology are adapted from these works, whereas the third-direction field is modified so that the configuration becomes a genuinely three-dimensional sheared-loop system rather than a simple extrusion of an earlier 2.5D model with a spatially uniform guide field. The magnetic field is initialized as
\begin{equation}
\begin{aligned}
B_x &= B_0\!\left[\cos(k_x x)e^{-k_x(z-z_0)}
      - \cos(3k_x x)e^{-3k_x(z-z_0)}\right], \\
B_y &= 0.3\,B_0\,e^{-k_x(z-z_0)}, \\
B_z &= B_0\!\left[-\sin(k_x x)e^{-k_x(z-z_0)}
      + \sin(3k_x x)e^{-3k_x(z-z_0)}\right].
\end{aligned}
\end{equation}
with $B_0 = 5$, $k_x = \pi/L$, $L = 10$, and $z_0 = -0.4$. The $x$--$z$ components form a superposition of a fundamental arcade and a third-harmonic arcade, producing the quadrupolar two-arcade topology inherited from earlier prominence models \citep{XiaChenKeppens2012,JercicKeppens2023,JercicJenkinsKeppens2024}, while the finite $y$ component introduces a shear field that decreases with height rather than acting as a constant guide field. This choice preserves the magnetic support required for evaporation-condensation condensation formation, while extending the inherited 2.5D configuration into a true three-dimensional sheared-loop system and avoiding excessively long $y$ excursions of field lines in regions where the in-plane field is weak.

The simulation is performed in a Cartesian domain spanning $[-5, 5] \times [-5, 5] \times [0, 5]$ in normalized units, where one length unit corresponds to $10~\mathrm{Mm}$. The simulation domain is discretized using a uniform $600 \times 600 \times 600$ grid, corresponding to a spatial resolution of $83.3~\mathrm{km}$. The initial atmosphere is prescribed as a gravitationally stratified chromosphere--transition-region--corona system following the general prominence-formation setup of \citet{XiaChenKeppens2012}. In such models, the lower atmosphere is represented by a cool chromosphere, while the coronal temperature stratification above the transition region is constructed from a conductive equilibrium rather than by ad hoc interpolation. The atmosphere is thermally and mechanically stratified in a physically consistent manner before prominence formation begins. Radiative losses are prescribed by the \texttt{Colgan\_DM} cooling curve, which combines the high-temperature tabulation of \citet{colgan_radiative_2008} with the low-temperature tabulation of \citet{dalgarno_heating_1972}.

Prior to the condensation stage, the atmosphere is first relaxed under a background heating term into a quasi-equilibrium containing a chromosphere, a thin transition region, and a hot low-$\beta$ corona.
To maintain the initial coronal equilibrium during the relaxation stage, we apply a background volumetric heating term of the form
\begin{equation}
    H_{\rm bg}(z) = H_{0}\exp\left(-\frac{z}{\lambda_{\rm bg}}\right),
\end{equation}
where \(H_{0}=10^{-4}\ {\rm erg\ cm^{-3}\ s^{-1}}\), and \(\lambda_{\rm bg}=50  \ {\rm Mm}\) is the heating scale height.

After this initial relaxation, an additional localized heating component is introduced near the transition region to drive chromospheric evaporation into the overlying coronal loops. Following the stochastic heating prescriptions of \citet{Zhou2020} and \citet{Li2022}, the heating function is generalized to three dimensions as
\begin{equation}
H_{\mathrm{loc}}(x,y,z,t) =
 H_{1}H_{xyz}(x,y,z)H_{z}(z)H_{t}(t) \,,
\end{equation}
where $H_{xyz}(x,y,z)$ represents the heating distribution constructed from a superposition of randomly phased Fourier modes in all the $x$, $y$, and $z$ directions. The mode amplitudes follow a power-law spectrum with an index of $5/6$. The heating is concentrated in the lower atmosphere and decreases exponentially above $z=2~\mathrm{Mm}$. Its temporal variation consists of a sequence of Gaussian heating episodes with durations centered at $5~\mathrm{min}$ and randomly varied by $\pm 75~\mathrm{s}$. The localized heating is switched on after a prescribed start time and is added to the energy equation with a characteristic amplitude of $H_{1}=0.1~\mathrm{erg~cm^{-3}~s^{-1}}$.

\begin{figure}
\centering
\includegraphics[width=1.05\columnwidth]{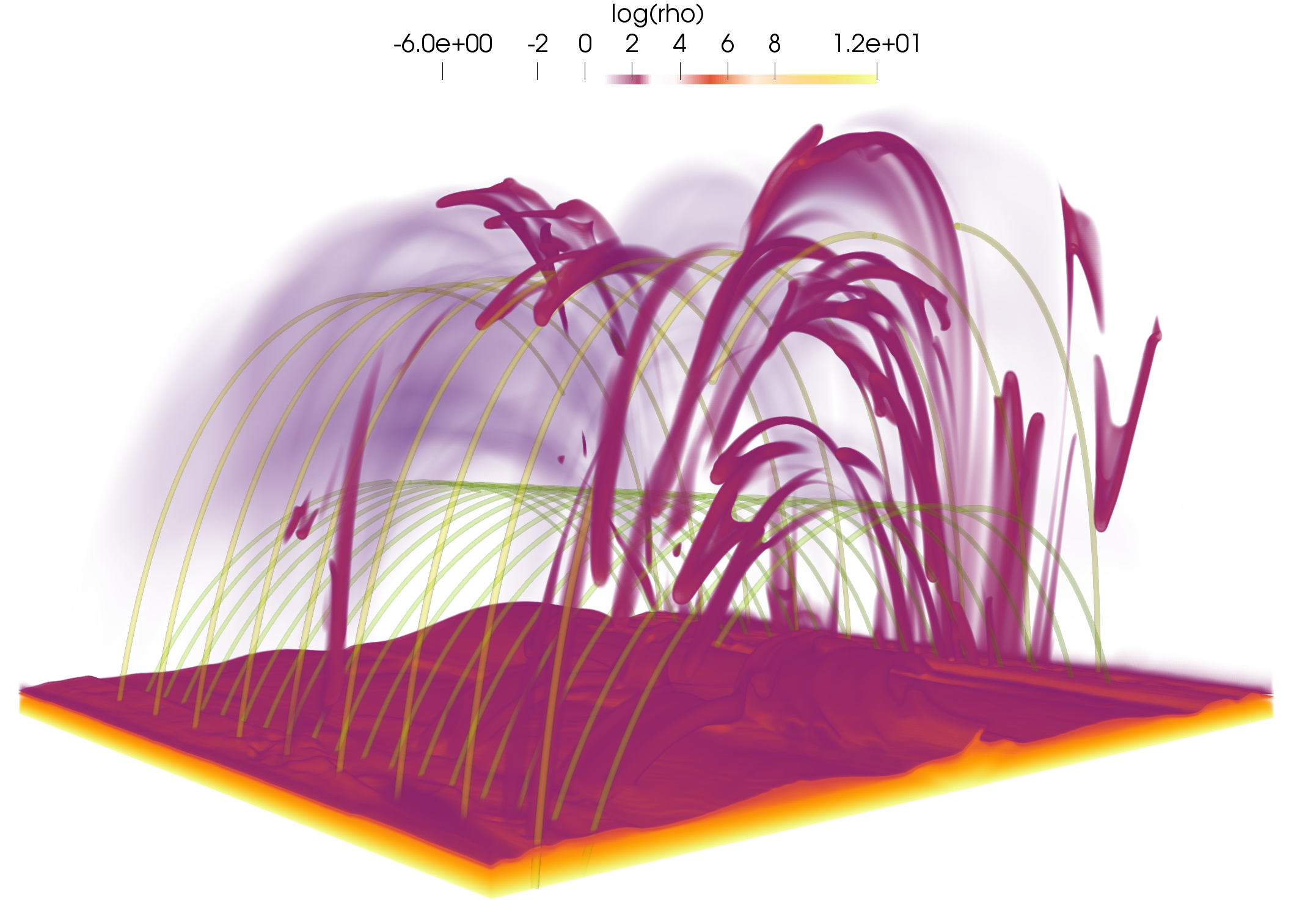}
\caption{Volume rendering of the condensed structures emerging in the ffhd simulation.  Filamentary overdensities continuously form and flow along field lines reminiscent of coronal rain.  A multi-peaked transfer function was selected to show near isocontours of density and we also show a sample of the frozen-in-time background magnetic field lines.  }
\label{fig:agile_ffhd_render}
\end{figure}

As evaporated mass accumulates in the coronal loops, optically thin radiative losses become increasingly important. Once local heating and field-aligned thermal conduction can no longer compensate for these losses, the plasma undergoes runaway cooling and condenses in situ. The present setup therefore follows the classical evaporation--condensation scenario, with thermal instability providing the immediate trigger for the formation of a multiphase plasma in which dense, cool condensations are embedded within the hot, tenuous corona. As shown in Figure~\ref{fig:agile_ffhd_render}, filamentary condensations form continuously and are subsequently channeled along the sheared quadrupolar magnetic field toward the lower atmosphere. Their evolution is therefore more characteristic of coronal rain than of magnetically supported prominence plasma. 

\begin{figure}
\centering
\includegraphics[width=\columnwidth]{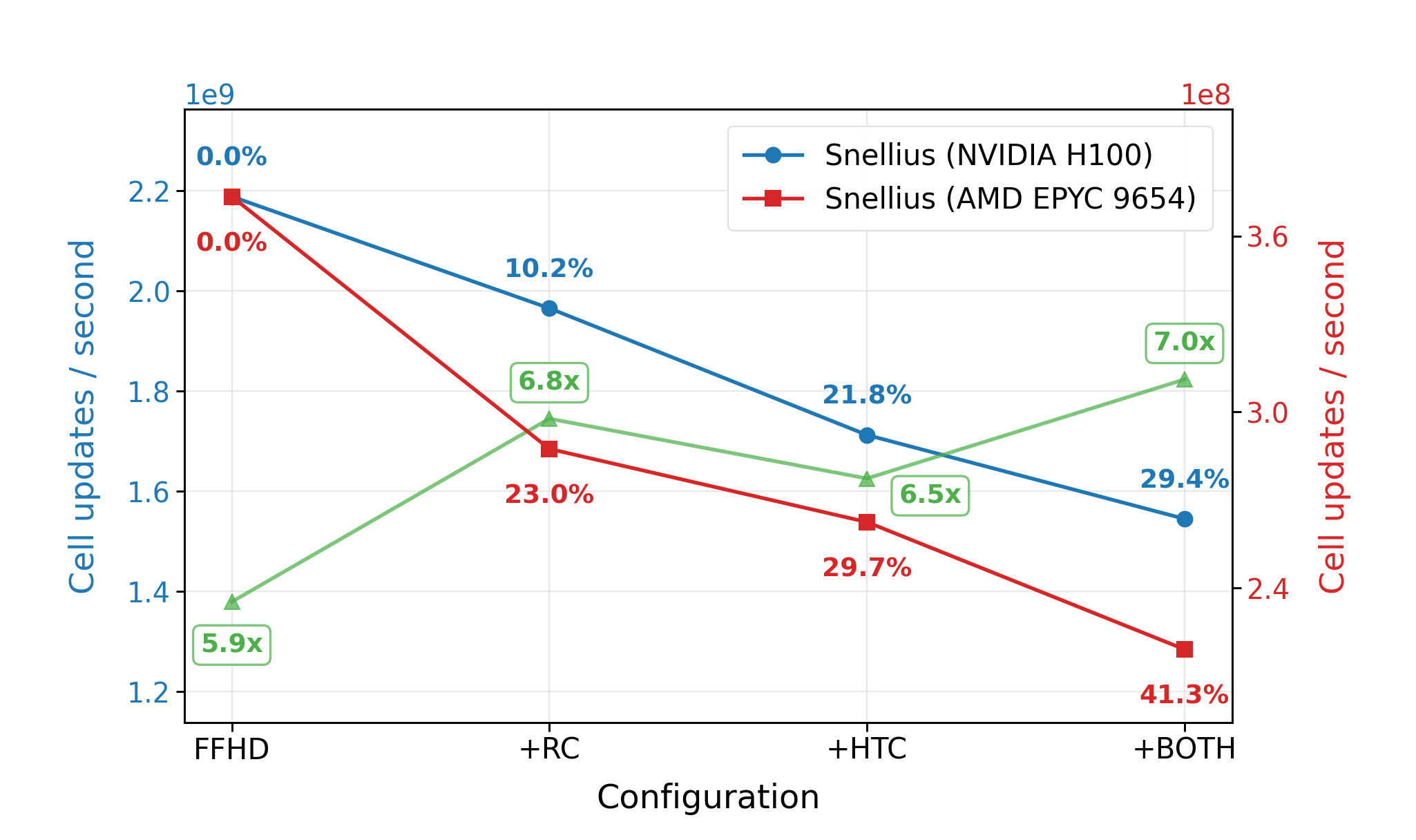}
\caption{Node-level performance comparison for the FFHD simulation case between running in \agile and \amrvac. Results are shown for the baseline FFHD configuration and for configurations including radiative cooling (\texttt{RC}), hyperbolic thermal conduction (\texttt{HTC}), and both source terms (\texttt{BOTH}). The blue and red curves show the throughput of \agile and \amrvac, respectively, measured in \texttt{Cell updates / second}, while the green curve gives the corresponding throughput ratio. The percentage labels indicate the throughput reduction relative to the baseline FFHD configuration for each implementation.
}
\label{fig:agile_ffhd_performance}
\end{figure}

Figure~\ref{fig:agile_ffhd_performance} complements the scientific results by quantifying the node-level computational performance of the FFHD simulation case and the additional cost introduced by the thermodynamic source terms. The comparison is performed between \agile running on one NVIDIA H100 GPU node and \amrvac running on one AMD EPYC 9654 CPU node on Snellius. Performance is measured in \texttt{Cell updates/second}, representing the total computational throughput achieved by each node.

Relative to the baseline FFHD configuration, the throughput decreases for both implementations when radiative cooling (\texttt{RC}), hyperbolic thermal conduction (\texttt{HTC}), and their combination (\texttt{BOTH}) are included. For \agile, the throughput is reduced by 10.2\%, 21.8\%, and 29.4\% for the \texttt{RC}, \texttt{HTC}, and \texttt{BOTH} configurations, respectively. The corresponding reductions for \amrvac are larger, reaching 23.0\%, 29.7\%, and 41.3\%. Consequently, the node-level AGILE-FFHD-to-AMRVAC throughput ratio increases from 5.9$\times$ for the baseline FFHD case to 6.8$\times$ with \texttt{RC}, remains at 6.5$\times$ with \texttt{HTC}, and reaches 7.0$\times$ when both source terms are included. These results demonstrate that the richer thermodynamic treatment introduces a measurable computational cost in both implementations, while \agile retains a substantial node-level throughput advantage, particularly for the most physics-complete configuration.

\subsection{Cooled Orszag-Tang evolution}\label{sec:otc}

As a demonstration of the MHD physics module, we introduce a 3D variant of an Orszag-Tang evolution. In the original 2D study by \cite{OT1979}, an incompressible MHD evolution on a doubly-periodic domain studied how a vortical flow with superposed double magnetic island structure behaved under visco-resistive settings. A more challenging version was pioneered by \cite{PD1991} where ideal, compressible MHD follows the same initial state, now quantified by its plasma beta $\beta$ and Mach number $M$. This latter test became a standard benchmark and is also used in \cite{Mignone2010,LesurBaghdadiEtAl2023}, while modified versions for two-fluid (plasma-neutral) settings \citep{Beatrice2022} or relativistic MHD \citep{Bart2008,2020ApJ...900..100R} have also been reported. The 2D compressible evolution for Mach $M=1$ and $\beta=3.33$ is most frequently studied, and is known to lead to a rich shock-dominated evolution demonstrating all MHD shock types \citep{Snow2021}.

The variant showcased for AGILE takes a triple-periodic domain of size $[0,1]^3$, sets the adiabatic
index $\hat{\gamma}= 1.6666667$, and takes parameters $\rho(t=0)=\rho_0=4$, $T(t=0)=T_0=0.25$ and $\beta_0=3$. The first two values quantify the initial uniform density and temperature, with the ideal gas law determining the uniform pressure $p_0=\rho_0T_0=1$, all in code units. The plasma beta determines the dimensionless magnetic field strength $B_0=\sqrt{2p_0/\beta_0}$. This enters the initial incompressible flow pattern and the solenoidal magnetic field as
\begin{eqnarray}
\mathbf{v}& = & \left(-\sin(2\pi y) \cos(2\pi z), \sin(2\pi x) \cos(2\pi z), 0\right) \,, \\
\mathbf{B}& = & B_0\left(-\sin(2\pi y) \cos(2\pi z), \sin(4\pi x) \cos(2\pi z), 0\right) \,. 
\end{eqnarray}
We use TVDLF with minmod limiting, but we add a user-defined uniform heating as well as optically thin radiative losses. The latter uses the Colgan-DM cooling curve (available cooling curves and references for them are in \cite{Joris2021}) and assigns physical units by using a fixed Helium abundance 0.1, and setting the (box size) unit length to 1 Mm, the code unit temperature to 1 MK, and a number density of $10^9 \mathrm{cm}^{-3}$. The time-independent, uniform background heating simply acts to balance the uniform, constant cooling obtained for the $t=0$ thermodynamic state exactly. The setting above has the initial Mach number reaching up to $M_0=1/\sqrt{\hat{\gamma}T_0}\approx 1.55$. The initial flow and magnetic field vectors have no $z$-components, but do show full 3D variations. 

We run till time $t=1.8$ and use 4 AMR levels such that the $100^3$ base resolution becomes effectively $800^3$. Our refinement is weighted on density and total energy, but the initial uniform thermodynamic state implies that the simulation starts with only grid level one activated, using 1000 blocks of $10^3$ size. From about $t\approx 0.4$ onwards, there are no base level blocks anymore, and the later evolution occasionally involves more than 400000 level 4 blocks. Since the simulation begins with few blocks, we start the simulation on 64 nodes of LUMI and restart with 256 nodes later on.  The run finished in 4 hours and 15 minutes wallclock time (of which one hour on 64 nodes). 

The evolution itself is showing the typical formation of current sheets due to the compressive deformation of the magnetic `islands', and fast mode shock fronts that separate off these compression sites. The addition of optically thin cooling induces runaway condensation associated with thermal instability, very similar to the FFHD scenario studied in Section~\ref{sec:ffhd}. 

\begin{figure*}
\centering
\includegraphics[width=\linewidth]{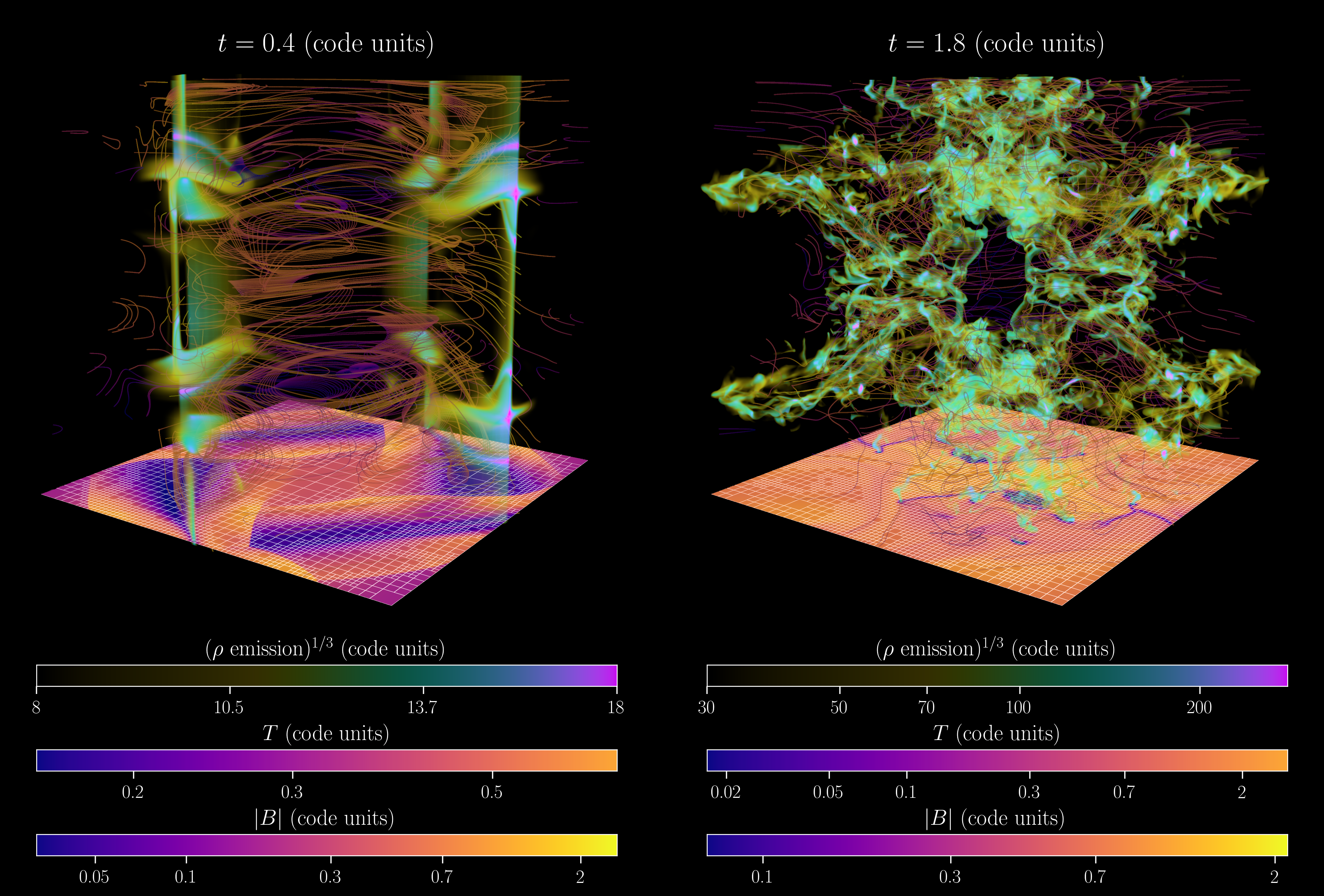}
\caption{
The cooled Orszag-Tang state at times $t=0.4$ and $t=1.8$, showing magnetic field lines and strength, a $T$ slice, and volumetrically regions of relatively high density indicating the formation of condensations due to cooling instability.
}
\label{fig:OTrender}
\end{figure*}

An impression of the early- and late-time states is shown in Figure \ref{fig:OTrender}.  At $t=0.4$, the evolution closely matches the known 2D case where magnetic tension starts acting against the vortical flow.  At $t=1.8$, the initial slab symmetry is lost and multiple low temperature fragmentation sites have developed.  The 2D slice identifies thin connected low temperature sheets which relate to high density condensations in the folds and edges of these sheets.  The figure was obtained from the native AMR simulation snapshot ``.dat'' file with the raytracing library Intel Embree\footnote{\url{https://www.embree.org/}}.  A big advantage of Embree is that large datasets can be processed (in this case a snapshot has a size of $\sim 30\,\rm GB$) on a laptop class computer since data is read ``on-demand'', obeying a fixed, user-specified RAM limit for the software.  

\subsection{Relativistic Jet}

As demonstration for the relativistic hydrodynamics module we here discuss a jet propagation simulation.  Powerful jets from active galactic nuclei, in particular FR-II type jets with powers $\gtrsim 10^{45}\rm erg\, s^{-1}$, are thought to remain relativistic on scales of $\sim 100\, \rm kpc$ until they decelerate via a moving shock complex at their jet head \citep{BlandfordRees1974,PorthKomissarov2015}.  The emerging flow structures, in particular shocks, shear and mixing regions are thought to play an important role in the acceleration of cosmic rays, the heating of the inter-cluster-medium (ICM) and the feedback between central engine and the galaxy at large \citep[see reviews by][]{BegelmanBlandfordEtAl1984,Fabian2012}.  Relativistic hydrodynamic simulations have widely been applied to investigate these processes \citep[e.g.][]{RossiMignoneEtAl2008,PeruchoQuilisEtAl2011,PeruchoMartiEtAl2018,vanderWesthuizenvanSoelenEtAl2019,SeoKangEtAl2021} and jet propagation is a common use case for such codes.  

Here we follow closely the setup by \cite{SeoKangEtAl2021} with the only change that we use ``reflective'' instead of ``outflow'' boundary conditions outside of the jet injection nozzle.  We also use the approximate Synge EOS.  The simulation corresponds to an underdense ($\eta:=\rho_j/\rho_b=10^{-5}$) jet injected with Lorentz factor of 7 into a domain with uniform density and pressure. The ambient medium is scaled to a number-density $n=10^{-3} \rm cm^{-3}$ with a temperature of $5\times 10^7\rm K$ -- representative of the hot ICM.  The radius of the jet inlet is scaled to $r_j=1000 \rm pc$ which sets a jet power of $1.09\times10^{45}\rm erg/s$.  From the expected advance speed of the jet head $v_{\rm head}$ \citep{MartiMuellerEtAl1997}, we obtain a jet crossing time of $t_{\rm cross} = r_j/v_{\rm head} \simeq 10^5\rm years$ and we simulate the evolution for $6.5\rm Myr$ or $42.8 t_{\rm cross}$.   
The jet inlet is resolved by $r_j/\Delta x = 12$ cells and we set up a cubic domain of $L_x = L_y = L_z = 80 r_{j}$ side length.  The simulation utilizes 4 AMR levels and the grid is automatically refined (based on gradients of density) as the jet propagates into the domain.  At the end of the simulation, 31296 computational blocks are activated (each containing $20^3$ grid cells), while the simulation started out with only 300.  The simulation uses \texttt{LF} fluxes and Van Leer spatial reconstruction, along with a threestep Runge-Kutta time integration and CFL factor of $0.4$.  No flooring or other robustness fallback is required for this test.  

\begin{figure*}
    \centering
    \includegraphics[width=1\linewidth]{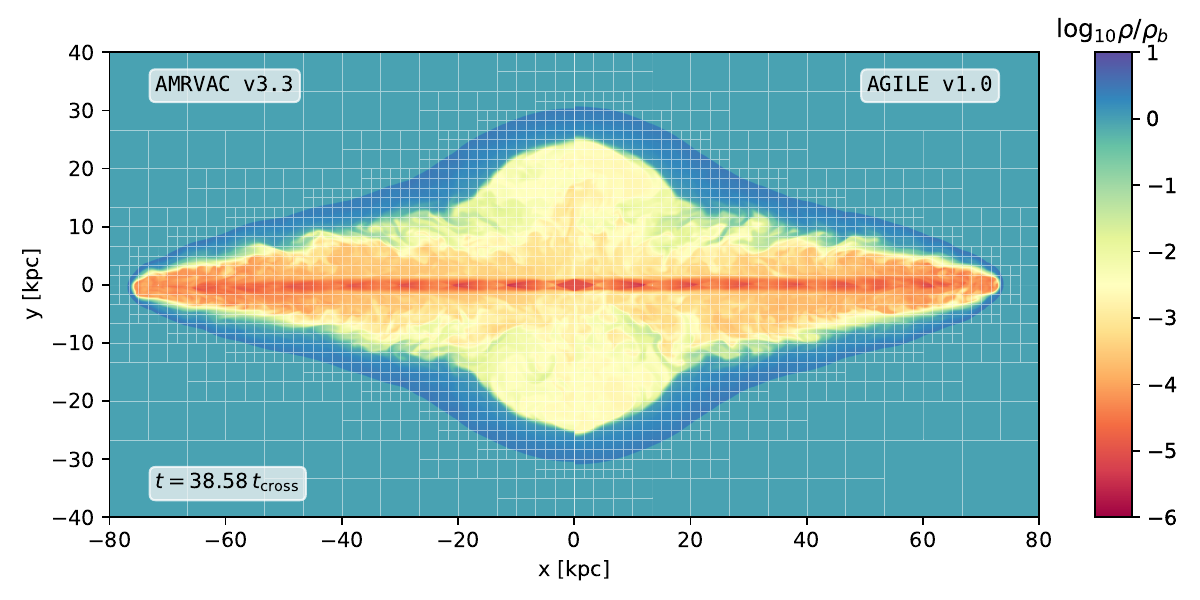}
    \caption{Propagation of a Lorentz factor $\Gamma=7$ relativistic jet simulation after $38$ crossing times.  Slice along jet axis comparing \amrvac (left) and \agile (right).  See \url{https://surfdrive.surf.nl/s/D7TewDy7ciNX5sk} for an animated version of the figure. }
    \label{fig:reljet}
\end{figure*}

As a means of validation, we have also ran the identical setup with \amrvac.  Since initial condition and diagnostic subroutines are largely identical between \agile and \amrvac, this required only minimal adjustments to the setup and parameter file. 
A comparison of the two simulations is shown in Figure \ref{fig:reljet} where the left-wards propagating jet shows the \amrvac solution and the right-wards propagating jet the \agile solution.  Both cases show very good qualitative agreement in the overall morphology and in particular the spacing of the repeated recollimation shocks is well matched between both codes.  

Comparing the performance and time to solution, the \agile job finished after $16\, \rm hours$ on a full Snellius node featuring 4 H100 GPUs, corresponding to $12\,288$ SBUs.  The average CUPS were $3.02\times 10^8$ per device resulting in a total of $1.2\times 10^9$ CUPS.\footnote{This performance figure excludes an overhead ($42\%$) for time spent on frequent slice-IO. The run produced 8000 2D slices and 200 3D restart snapshots of modest size (below 16 GB per snapshot file).}.  The \amrvac run on the other hand was executed on 512 EPYC Rome cores (four Snellius nodes) with an average performance of $2.6\times 10^5$ CUPS per core and thus a total performance of $1.3\times 10^8$ CUPS.  The \amrvac run completed after a wallclock time of $85.6\, \rm hours$, taking $5.3$ times longer than \agile.  The cost to solution was $43\,827$ SBU rendering the \agile run $3.6$ times cheaper (considering the full runtime including IO).  

\section{Discussion and Conclusions}\label{sec:conclusions}

We have presented the first version of \agile, a highly performant and scalable GPU accelerated adaptive mesh refinement framework for the solution of near- conservation laws.  As main physics modules, we target compressible (magneto-) hydrodynamics which is applicable in numerous astro- and solar-physical contexts.  \agile inherits its mesh handling and several physics routines (e.g. the radiative cooling module) from \amrvac to which it remains largely compatible.  Initial conditions can be shared nearly 1:1 between the codes and the datafiles generated remain 100\% exchangeable.  This allows \agile to take advantage of a plethora of science ready setups and postprocessing workflows.  In fact, the \texttt{tests} folder in \amrvac \texttt{v3.3} comprises more than 100 hd and mhd setups alone.  Regarding postprocessing and data analysis, on the one hand, \agile can do light processing on the fly, for example performing dimensional reduction via slices, ``column densities'' and global integrals over the grid.  Furthermore, \agile's output can be read directly with standard postprocessing tools such as ParaView and VisIt (via \texttt{.vtu} files, see Figures \ref{fig:DoubleMach},  \ref{fig:agile_ffhd_render}) and yt (which reads native \texttt{.dat} snaphshot files, e.g. Figure \ref{fig:TRML_slices}).  Python tools to read native \texttt{.dat} files as well as \texttt{.vtu} files are also provided in \agile's repository (e.g. Figures \ref{fig:cc-cloud_3d} and \ref{fig:reljet} which were created with matplotlib).  A novel addition to the postprocessing workflows is native \texttt{.dat} file reading and highly customizable analysis utilizing ultimately the raytracing library Intel Embree (see Figure \ref{fig:OTrender}).  By reading data ``on-demand'' and limiting the RAM usage to a fixed, user-specified amount, Embree is able to render arbitrarily large datasets. Naturally, this property is very attractive for large scale simulations with snapshots that don't fit into the system memory of a typical workstation or compute node.

The performance-critical parts of \agile have been completely redesigned to minimize kernel launch overhead and global memory usage.  To this end, \agile uses a macro-kernel which parallelizes over the block loop and uses the vector units for all operations needed for one Runge-Kutta substep: 1. conversion to primitive variables, 2. reconstruction of interface values, 3. flux computation via Riemann solvers, 4. application of source terms and 5. update of the conserved variables.  

One potential concern of this approach is that performance degradation could occur when this macro-kernel runs out of registers on the SMP.  However, our detailed benchmarks have not yet indicated any extraordinary drop in performance when more complex physics is incorporated.  For example, in the coronal rain scenario discussed in Section \ref{sec:rain}, when incorporating both radiative cooling (which implies table lookups) and thermal conduction, we obtain a $\simeq29\%$ drop in performance -- compared to a $\simeq43\%$ performance loss in the CPU code \amrvac.  
Also the special relativistic hydrodynamics module which incurs an iterative (therefore branching) Newton-Raphson ``con2prim'' solver showed good performance without any further GPU specific optimizations (with an average of $3\times 10^8$ CUPS per H100 device in a full multi-GPU AMR simulation).  

Another performance critical redesign concerns the ghost-cell exchange.  \agile implements optimally sized dynamic buffers which minimize memory overhead and the number of MPI messages between neighboring MPI ranks.  \agile provides implementations for both:  direct memory access between GPUs (aka GPUDirect) and a host-copy fallback option.  

With these design choices, \agile obtains good performance even with block sizes of $16^3$ cells or smaller and shows excellent strong scaling on GPUs and CPUs alike.  
To stress-test \agile's AMR handling, we have shown a six-level double Mach reflection application (Section \ref{sec:wc}) which has evolved up to $10^6$ dynamical blocks (each with only $12^3$ cells) on 2048 GPUs demonstrating the feasibility of deeply nested AMR simulations on GPUs.  

The single GPU benchmarks show that performance is robust for a range of problem- and block-sizes and the performance on a range of devices is generally ordered according to the memory transfer speed of the devices.  
It is fortuitous that the algorithms employed within \agile are memory bound (see Appendix \ref{sec:roofline}): \agile will likely benefit from future GPU generations that are optimized for AI applications which are likewise often memory bound.  

On the most recent hardware available at the time of writing (Nvidia B200), for a simple hydrodynamics benchmark, we have obtained a peak performance of $2\times 10^9$ cell updates per second (CUPS) per device.  While care must be taken when comparing performance metrics across non-standardized test cases and facilities, it seems that \agile is at least on-par with other recent codes such as \athenaK \citep[see][which demonstrates $10^9$ CUPS on the GH200 -- likely for the full RK2 step though]{StoneMullenEtAl2026} and \gpluto \citep[see performance per node for an MHD case in][]{RossazzaMignoneEtAl2025}.  

\agile's performance is also consistent with \foap which has been designed from the ground up to maximize performance on GPUs. For example, in addition to the macro-kernel update, \foap employs a rather complex non-branching asynchronous ghostcell exchange algorithm which should be ideal for threaded GPU operations.  Comparing the performance, for the case where \agile performs best ($512^3$ domain, blocksize of $32^3$) the corresponding hydrodynamic uniform grid benchmark discussed in \citet{TeunissenFOAP} reaches $1.45\times 10^9$ CUPS whereas \agile obtains $1.3\times 10^9$ CUPS on identical hardware (H100 SXM5 96GB).  For \foap, the peak performance is reached for the same problem size but bigger blocks of $64^3$ cells at $1.52\times 10^9$ CUPS.  It is reasonable to assume that these differences of order $\sim15\%$ are explained by different algorithmic choices in the ghost-cell filling part.  Thus while \agile could still be sped up by $\sim 15\%$, this would go on the account of simplicity, maintainability and consistency with \amrvac and we do not consider it worth the effort at this point.  

Some remarks about our experience with OpenACC.  Two important factors in our decision for OpenACC were performance portability and the ability to write kernel functions in native fortran.  Via the NVIDIA HPC package and the HPE/Cray compiler, \agile indeed runs on both NVIDIA and AMD GPUs.  However, getting the code to perform well with both toolchains incurred significant work.  Without dedicated software engineers from the Netherlands eScience center in our project, we would surely have given up on LUMI.  

The main issue with porting \agile to LUMI was the sometimes differing interpretation of the OpenACC standard by the vendors.  Email contact with HPE and the clear debugging output of the Cray compiler did help to clarify such instances. In one extreme case, we needed to rename variables since they were aliasing on the device due to the lack of a module scoping unit with the installed Cray compiler.  
Other problems we encountered with the OpenACC implementations (both with NVIDIA and HPE/Cray) were the lack of procedure pointers on the device, somewhat ``flaky'' support for nested structures with pointer or allocatable components (regarding transfer and update of the components, e.g. when the memory footprint changes on host -- this often resulted in memory access faults on the device) and performance degradation when kernels call functions from other modules.  

We have found workarounds for all essential cases and have documented the discovered compatibility issues on this website: \url{https://v1kko.github.io/OpenACC_behaviour/}, hoping that this will help other practitioners and vendors alike to improve and unify the OpenACC ecosystem.
Once we found a ``supported featureset'', developing and extending \agile became quite straight-forward however.  Expressing the algorithms as annotated multi-dimensional ``collapsed loops'' typically immediately gives good performance, and the fact that everything can be coded in native simple fortran means that \agile can now be easily extended by students.

Still, for full performance portability including intel and general AMD hardware, adding another decorator layer of openMP offloading will be required.   Work towards this end has begun in the \agile repository and is already complete for the \foap code via an abstraction layer.  

The presented \agile\texttt{v1.0} is already a capable and cost-efficient GPU framework with demonstrated reduced cost to solution of up to a factor of $5$ compared to the existing CPU implementation in \amrvac.  We currently carry out simulations to study the physics of the solar corona and of relativistic jets.  These applications will be discussed in forthcoming papers.  
In the future, besides porting of established physics modules, reconstructions and time-stepping schemes from \amrvac, we plan to continuously update the framework itself. This will start with the support for non-cartesian grids and the merge of the general relativistic physics modules from \bhac.  Future versions should also include constrained transport and will improve parallel IO which can quickly become the bottleneck for large scale simulations.

\section*{Acknowledgements}

This work was supported by the Netherlands eScience Center under grant number NLESC.OEC.2023.008 (AGILE).
This work has made use of resources and expertise provided by SURF Experimental Technologies Platform, which is part of the SURF cooperative in the Netherlands, under project no. SURF-ETP0037.
We acknowledge the Dutch Research Council (NWO) in The Netherlands for awarding this project access to the LUMI supercomputer, owned by the EuroHPC Joint Undertaking, hosted by CSC (Finland) and the LUMI consortium through the ‘Computing Time on National Computer Facilities’ call.  
This work was further supported by the Dutch national e-infrastructure with the support of SURF Co-operative, project EINF-11051. 

RK, HW, AK and OW acknowledge funding from the KU Leuven C1 project C16/24/010 UnderRadioSun and the Research Foundation Flanders FWO project G0B9923N Helioskill, and computational resources and services provided by the VSC (Flemish Supercomputer Center), funded by the Research Foundation Flanders (FWO) and the Flemish Government, department EWI.
JV acknowledges support from Research Foundation -- Flanders (FWO) postdoctoral fellowship 1255226N and KU Leuven postdoctoral mandate PDMT1/24/012.
HO is supported by the Individual CEEC program -- 5th edition funded by
the Portuguese Foundation for Science and Technology (FCT, \url{https://ror.org/00snfqn58}). This work is
supported by CIDMA (\url{https://ror.org/05pm2mw36}) under the
Portuguese FCT, Grants UID/04106/2025 (\url{https://doi.org/10.54499/UID/04106/2025}) and UID/PRR/04106/2025 (\url{https://doi.org/10.54499/UID/PRR/04106/2025}).
The authors acknowledge support from the projects
CERN/FIS-PAR/0024/2021 and 2022.04560.PTDC
and the European Horizon Europe staff exchange
(SE) programme HORIZON-MSCA2021-SE-01 grant NewFunFiCO (No. 10108625).

\textit{Software:}
          \agile (this paper), 
          \amrvac \citep{PorthXia2014,XiaTeunissenEtAl2018a,KeppensTeunissenEtAl2021,KeppensPopescuBraileanuEtAl2023}, 
          Intel Embree \citep{WaldWoopEtAl2014},
          yt \citep{yt},
          Python \citep{Python2007},
          Matplotlib \citep{Hunter2007, matplotlibv2},
          ParaView \citep{ParaView}
\section*{Data Availability}

A reproduction package (setups and parameter files) for all simulations discussed in this paper will be made available upon acceptance of the article.  The reproduction package can be found on the repository 
\url{https://doi.org/10.5281/zenodo.21392983}.



\bibliographystyle{rasti}
\bibliography{agile}




\appendix

\section{Roofline analysis}\label{sec:roofline}
In order to assess the performance of AGILE with respect to the peak performance of a GPU, we need to consider both the compute performance and memory bandwidth of a GPU. We do this through a roofline analysis. For this we consider the AGILE macro-kernel, which takes the majority of the compute time. The same AGILE setup as in Section~\ref{sec:singleGPU} is used, with a block size of $16^3$ and a domain size of $256^3$. We then run AGILE with either rocprof (AMD) or Nsight compute (NVIDIA) to obtain the relevant performance counters. Both tools directly give the values needed for the roofline: the performance in $10^9$ operations per second (GFLOP/s) and the arithmetic intensity in number of operations per byte transferred to/from global device memory (FLOPs/byte).

\begin{figure}
    \centering
    \includegraphics[width=\linewidth]{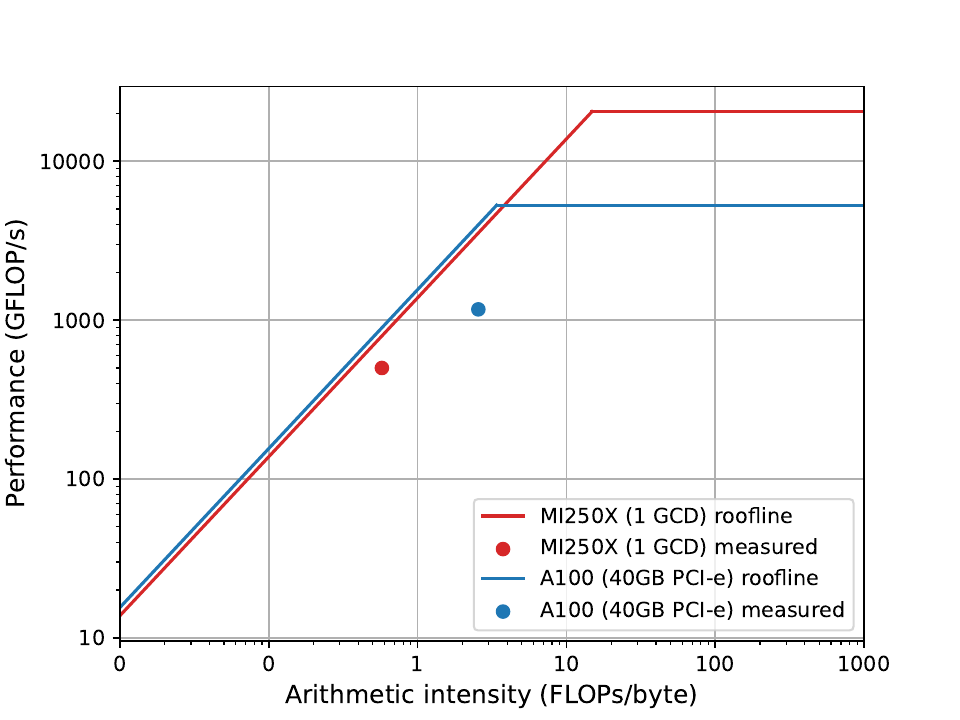}
    \caption{Roofline analysis of the AGILE macro-kernel. On the MI250X AGILE is memory bound, on the A100 it is still memory bound but close to the ridge point. The difference in arithmetic intensity is expected to be mainly due to the much larger L2 cache of the A100, leading to less global device memory traffic and thus a higher AI.}
    \label{fig:roofline}
\end{figure}

For AMD, we run on a single MI250X GCD. For NVIDIA, priviliged access to the GPU is required, which is not available on most HPC systems. We therefore run on a local system where this access is available. This system has a single A100 40G PCI-e GPU, which is comparable to the A100 in Snellius but in a different form factor with a lower power limit. The results are shown in Fig.~\ref{fig:roofline}.

On both GPUs, AGILE is memory bound, although it is close to the ridge point on the A100. The absolute performance is better on the A100, as was already evident from the amount of cell updates per second in Section~\ref{sec:singleGPU}. The peak FP64 performance of the MI250X is roughly four times higher than the A100, however this has no impact on the performance because the limiting factor is memory bandwidth, which is similar for both cards.

Interestingly, the arithmetic intensity is very different for both GPUs: on the A100 it is ~4.4 times larger than on the MI250X. There may be small differences due to different compiler optimizations, however the main difference is likely from the differences in cache sizes: The arithmetic intensity is determined by comparing the compute operations to the amount of data transferred to and from the global device memory. AGILE makes very efficient use of the GPU caches, achieving high hit rates and throughput. The L2 cache of the A100 is five times larger than that of the MI2520X GCD. With perfect cache use, this could lead to fives times less traffic from global device memory and hence a five times larger arithmetic intensity at the same number of computation operations.

\section{Domain and block size sweeps}\label{sec:singleGPUdetails}

The full results of the single device experiments from Section \ref{sec:singleGPU} are provided in Table \ref{tab:1GPUdetails}.  In case of many small blocks per device (32768), several experiments failed due to out of memory error (indicated as OOM) due to the overhead from block handling on the device.  The CPU tests have been performed with pure MPI parallelization.

\begin{table}
 \caption{Performance on single GPU/single CPU nodes, uniform grid 3D hydrodynamics.}
 \label{tab:1GPUdetails}
 \begin{tabular*}{\columnwidth}{p{1.5cm}llp{1cm}p{1.3cm}}
  \hline
  Device name (Bandwidth) & Problem size & Block size & Blocks per MPI task & Performance\newline [$10^8$ CUPS]\\
  \hline
\multirow[t]{9}{1.5cm}{\parbox[t]{1.5cm}{B200\\ (8 TB/s)}}
       & $128^3$ & $16^3$  & 512 & $12.2$ \\
       &         & $32^3$  & 64 & $3.04$ \\
       &         & $64^3$  & 8 & $0.43$ \\
       \cline{2-5}
       & $256^3$ & $16^3$  & 4096 & $15.52$ \\
       &         & $32^3$  & 512 & $16.24$ \\
       &         & $64^3$  & 64 & $3.37$ \\
       \cline{2-5}
       & $512^3$ & $16^3$  & 32768 & $16.30$ \\
       &         & $32^3$  & 4096 & $19.91$ \\
       &         & $64^3$  & 512 & $18.10$ \\
       \hline
\multirow[t]{9}{1.5cm}{\parbox[t]{1.5cm}{GH200\\ (4.8 TB/s per H200)}}
       & $128^3$ & $16^3$  & 512 & $8.4$ \\
       &         & $32^3$  & 64 & $3.13$ \\
       &         & $64^3$  & 8 & $0.44$ \\
       \cline{2-5}
       & $256^3$ & $16^3$  & 4096 & $11.14$ \\
       &         & $32^3$  & 512 & $10.72$ \\
       &         & $64^3$  & 64 & $3.46$ \\
       \cline{2-5}
       & $512^3$ & $16^3$  & 32768 & $10.70$ \\
       &         & $32^3$  & 4096 & $13.98$ \\
       &         & $64^3$  & 512 & $11.65$ \\
       \hline
\multirow[t]{9}{1.5cm}{\parbox[t]{1.5cm}{H100\\SXM5 96GB\\ (3.35 TB/s)}}
       & $128^3$ & $16^3$  & 512 & $7.94$ \\
       &         & $32^3$  & 64 & $3.10$ \\
       &         & $64^3$  & 8 & $0.44$ \\
       \cline{2-5}
       & $256^3$ & $16^3$  & 4096 & $10.32$ \\
       &         & $32^3$  & 512 & $10.05$ \\
       &         & $64^3$  & 64 & $3.39$ \\
       \cline{2-5}
       & $512^3$ & $16^3$  & 32768 & $10.86$ \\
       &         & $32^3$  & 4096 & $13.01$ \\
       &         & $64^3$  & 512 & $11.08$ \\
       \hline
\multirow[t]{9}{*}{\parbox[t]{1.5cm}{A100\\ SXM4 80GB\\ "wICE"\\ (2.04 TB/s)}}
& $128^3$ & $16^3$  & 512 & $5.06$ \\
       &         & $32^3$  & 64 & $2.58$ \\
       &         & $64^3$  & 8 & $0.37$ \\
       \cline{2-5}
       & $256^3$ & $16^3$  & 4096 & $6.45$ \\
       &         & $32^3$  & 512 & $6.67$ \\
       &         & $64^3$  & 64 & $2.86$ \\
       \cline{2-5}
       & $512^3$ & $16^3$  & 32768 & $6.70$ \\
       &         & $32^3$  & 4096 & $8.28$ \\
       &         & $64^3$  & 512 & $7.39$ \\
       \hline
\multirow[t]{9}{*}{\parbox[t]{1.5cm}{A100\\ SXM4 40GB\\ "Snellius"\\ (1.55 TB/s)}}
& $128^3$ & $16^3$  & 512 & $4.82$ \\
       &         & $32^3$  & 64 & $2.39$ \\
       &         & $64^3$  & 8 & $0.34$ \\
       \cline{2-5}
       & $256^3$ & $16^3$  & 4096 & $6.33$ \\
       &         & $32^3$  & 512 & $6.25$ \\
       &         & $64^3$  & 64 & $2.67$ \\
       \cline{2-5}
       & $512^3$ & $16^3$  & 32768 & OOM \\
       &         & $32^3$  & 4096 & $7.89$ \\
       &         & $64^3$  & 512 & $6.93$ \\
       \hline
\multirow[t]{9}{*}{\parbox[t]{1.5cm}{ MI250X\\ 1 GCD\\ "LUMI"\\ (1.6 TB/s per GCD)} }
& $128^3$ & $16^3$  & 512 & $1.66$ \\
       &         & $32^3$  & 64 & $1.54$ \\
       &         & $64^3$  & 8 & $0.35$ \\
       \cline{2-5}
       & $256^3$ & $16^3$  & 4096 & $1.88$ \\
       &         & $32^3$  & 512 & $2.16$ \\
       &         & $64^3$  & 64 & $1.80$ \\
       \cline{2-5}
       & $512^3$ & $16^3$  & 32768 & OOM \\
       &         & $32^3$  & 4096 & $2.16$ \\
       &         & $64^3$  & 512 & $2.23$ \\
       \hline
\multirow[t]{9}{*}{\parbox[t]{1.5cm}{ L40S\\ (0.86 TB/s)} }
& $128^3$ & $16^3$  & 512 & $2.00$ \\
       &         & $32^3$  & 64 & $1.47$ \\
       &         & $64^3$  & 8 & $0.19$ \\
       \cline{2-5}
       & $256^3$ & $16^3$  & 4096 & $2.94$ \\
       &         & $32^3$  & 512 & $2.18$ \\
       &         & $64^3$  & 64 & $1.53$ \\
       \cline{2-5}
       & $512^3$ & $16^3$  & 32768 & OOM \\
       &         & $32^3$  & 4096 & $3.26$ \\
       &         & $64^3$  & 512 & $2.60$ \\
 \end{tabular*}
\end{table}

\begin{table}
 \begin{tabular*}{\columnwidth}{p{1.5cm}llp{1cm}p{1.3cm}}
       \hline
\multirow[t]{6}{*}{\parbox[t]{1.5cm}{ TitanV\\ (0.65 TB/s)} }
& $128^3$ & $16^3$  & 512 & $1.87$ \\
       &         & $32^3$  & 64 & $2.06$ \\
       &         & $64^3$  & 8 & $0.34$ \\
       \cline{2-5}
       & $256^3$ & $16^3$  & 4096 & $2.20$ \\
       &         & $32^3$  & 512 & $2.11$ \\
       &         & $64^3$  & 64 & $2.35$ \\
       \hline       
\multirow[t]{5}{*}{\parbox[t]{1.5cm}{\raggedright Dual EPYC\\ 9654 Genoa\\ (192 MPI tasks)}}
       & $128^3$ & $16^3$  & 2.66 & $1.99$ \\
       \cline{2-5}
       & $256^3$ & $16^3$  & 21.33 & $1.97$ \\
       &         & $32^3$  & 2.66 & $1.97$ \\
       \cline{2-5}
       & $512^3$ & $16^3$  & 170.66 & $2.02$ \\
       &         & $32^3$  & 21.33 & $2.51$ \\
       &         & $64^3$  & 2.66 & $1.97$ \\
  \hline
\multirow[t]{5}{*}{\parbox[t]{1.5cm}{\raggedright Dual EPYC\\ 7763 Milan\\ (128 MPI tasks)} }
& $128^3$ & $16^3$  & 4 & $3.54$ \\
       \cline{2-5}
       & $256^3$ & $16^3$  & 32 & $2.91$ \\
       &         & $32^3$  & 4 & $3.30$ \\
       \cline{2-5}
       & $512^3$ & $16^3$  & 256 & $2.96$ \\
       &         & $32^3$  & 32 & $3.48$ \\
       &         & $64^3$  & 4 & $3.48$ \\
\hline
 \end{tabular*}
\end{table}

\bsp	
\label{lastpage}
\end{document}